\documentclass[aps,pra,twocolumn,superscriptaddress]{revtex4}
\usepackage{amsfonts,amssymb,graphicx}
\usepackage[usenames,dvipsnames]{xcolor}
\usepackage{pdfpages}
\usepackage{footmisc}
\usepackage[normalem]{ulem}
\usepackage{color}
\begin{document}

\title{Longitudinal magnetization dynamics in the quantum Ising ring: A Pfaffian method based on correspondence between momentum space and real space}
% based on the real-space and momentum-space correspondence}
\author{Ning Wu}
\email{wun1985@gmail.com}
\affiliation{Center for Quantum Technology Research, School of Physics, Beijing Institute of Technology, Beijing 100081, China}
\begin{abstract}
\par As perhaps the most studied paradigm for a quantum phase transition, the periodic quantum Ising chain is exactly solvable via the Jordan-Wigner transformation followed by a Fourier transform that diagonalizes the model in the momentum space of spinless fermions. Although the above procedures are well-known, there remain some subtle points to be clarified regarding the correspondence between the real-space and momentum-space representations of the \emph{finite-size} quantum Ising ring, especially those related to fermion parities. In this work, we establish the relationship between the two fully aligned ferromagnetic states in real space and the two degenerate momentum-space ground states of the classical Ising ring, with the former being a special case of the factorized ground states of the more general XYZ model on the frustration-free hypersurface. Based on this observation, we then provide a Pfaffian formula for calculating real-time dynamics of the parity-breaking longitudinal magnetization with the system initially prepared in one of the two ferromagnetic states and under translationally invariant drivings. The formalism is shown to be applicable to \emph{large} systems with the help of online programs for the numerical computation of the Pfaffian, thus providing an efficient method to numerically study, for example, the emergence of discrete time crystals in related systems.
\end{abstract}

\maketitle

\section{Introduction}
\par The one-dimensional quantum Ising model with periodic boundary conditions (or a quantum Ising ring) is usually regarded as one of the two prototypical models for understanding quantum phase transitions~\cite{Sachdev}. Because of its exact solvability, the model has served as a testbed for a wide variety of physical phenomena including entanglement in quantum critical phenomena~\cite{pra2002,nature2002,vidal2003,pra2010}, quantum quenches across a quantum critical point~\cite{Zoller2005,Dzig2005,Sun2006,compo2012,wu2015}, dynamical quantum phase transitions~\cite{Heyl2013,Gong2019}, and more recently proposed discrete time crystals~\cite{Yu2019}, and so on.
\par It is well known that the quantum Ising ring can be solved by first performing the Jordan-Wigner transformations that convert the spin operators into spinless fermions, followed by a Fourier transform which maps the such obtained quadratic fermion model into a free fermion model in the momentum space~\cite{Lieb1962}. The presence of the boundary term connecting the last site of the spin chain to the first one does not lead to a simple cyclic structure in the Jordan-Wigner fermion representation. Consequently, the full Hilbert space is divided into two subspaces containing even and odd numbers of fermions, respectively. Although for large enough systems one can neglect the boundary corrections in the discussion of quantum phase transitions, it is both interesting and important to study the case of \emph{finite-size} systems in which the boundary terms and the resultant fermion-parity effects cannot be neglected~\cite{finiteising,Facc2009,Sire2010,Rams2014,Nishimori2015,Dong,Oshikawa2019}. For example, it is shown by different approaches that the true ground state of a finite-size quantum Ising ring with even number of sites lies in the subspace with even number of fermions~\cite{Sire2010,Rams2014,Nishimori2015,Katsura2012}.
\par It is also known that the quantum Ising chain with open boundaries can be mapped into Kitaev's $p$-wave superconductor chain with equal pairing and hopping strength~\cite{Kitaev2001,Fendley2012,Greiter2014}, which can be viewed as a special case of the correspondence between the more general open XYZ spin chain and the interacting Kitaev chain with open boundaries~\cite{Katsura2015}. For the open XYZ chain, there exists a frustration-free hypersurface in the parameter space on which the ground state is twofold degenerate and admits separable forms in the spin representation~\cite{Krumann,Muller1985,Cerezo2015}. Due to the equivalence between the XYZ chain and the interacting Kitaev chain, similar twofold degenerate separable fermionic ground states occur as well under the frustration-free condition~\cite{Katsura2015}. In general, the above two separable fermion ground states are neither orthogonal to each other nor have definite fermion parity. Nevertheless, simple equally weighted linear superpositions of the two yield two orthogonal states with distinct fermion parities~\cite{Katsura2015}. It was further shown in~\cite{Katsura2017} that the aforementioned two orthogonal states, one of which having odd fermion parity and the other having even parity, are also the ground states of the interacting Kitaev chain under periodic and antiperiodic boundary conditions, respectively.
\par Recently, nonequilibrium dynamics of a quantum Ising chain initialized in one of the ferromagnetic states has been considered. In Ref.~\cite{Gong2019}, the dynamics of the nearest-neighbor equal-time longitudinal correlation function under a quench starting with the ferromagnetic state is calculated and shown to exhibit nonanalytic signature that indicates a dynamical critical point. A simple scheme to generate a discrete time crystal was proposed in Ref.~\cite{Yu2019} by periodically kicking a quantum Ising chain initially prepared in one of the two ferromagnetic states, where the time evolution of the longitudinal magnetization was numerically calculated using exact diagonalization for relatively small system sizes.
\par In this work, we first show that the two momentum-space degenerate ground states of the periodic classical Ising model, which have definite fermion parities and are of the BCS form, are actually equivalent to the two equally weighted linear superpositions of the two ferromagnetic states in the spin representation. This correspondence is shown to be the special case of the XYZ chain under the frustration-free condition. Based on this observation and assuming that the system is prepared in one of the two ferromagnetic states, we are able to obtain the time-evolved state in the momentum space under \emph{translationally invariant drivings}. Similar to the initial state, the time-evolved state is also a linear combination of two states with distinct fermion parities, which makes the calculation of expectation values of \emph{parity-breaking} observables such as the longitudinal magnetization subtle due to the mixing of the two parity components. Nevertheless, by writing the BCS-type time-evolved state for each mode as the occupation state of two effective (time-dependent) Bogoliubov fermions, we are able to derive a Pfaffian formula for the expectation value of the longitudinal magnetization. With the help of free online algorithms that can achieve efficient numerical computation of the Pfaffian, the dynamics of the longitudinal magnetization can be probed for large systems and at long-time scales. As an illustration of the above formalism, we perform numerical simulations for sudden quenches and periodic delta kicks in the quantum Ising ring prepared in the ferromagnetic state.
\par The rest of the paper is organized as follows. In Sec.~\ref{secII}, we introduce the Hamiltonian of the quantum Ising ring and describe its diagonalization in detail. In Sec.~\ref{III}, we identify the two sub-ground states in each fermion parity sector and establish the relationship between the ferromagnetic states and the momentum-space BCS states of the classical Ising model. In Sec.~\ref{IV}, we derive the Pfaffian formula for calculating the dynamics of the longitudinal magnetization and discuss its application to the sudden quench and delta kicks. Conclusions are drawn in Sec.~\ref{V}.
\section{Model and diagonalization}\label{secII}
\par For the sake of completeness and to introduce the notations that will be used later, we provide in this section the details of the diagonalization of the quantum Ising ring.
\subsection{The Bogoliubov-de Gennes form}
\par The one-dimensional ferromagnetic quantum Ising model with $N$ sites is described by the Hamiltonian
\begin{eqnarray}\label{HQIM}
H_{\mathrm{QIM}}=-\sum^N_{j=1}(\sigma^x_j\sigma^x_{j+1}+g\sigma^z_j),
\end{eqnarray}
where $\sigma^\alpha_j$ ($\alpha=x,y,z$) are the Pauli matrices on site $j$, $g\geq 0$ is the transverse field along the $z$-direction, and we have set the nearest-neighboring exchange interaction to unit for simplicity. We consider \emph{even} $N$ and impose periodic boundary conditions, i.e., $\vec{\sigma}_{N+1}=\vec{\sigma}_1$. The model exhibits a quantum phase transition at the critical field $g_c=1$ between the ordered phase for $0\leq g<1$ and disordered phase for $g>1$.
\par To diagonalize $H_{\mathrm{QIM}}$, we first perform the Jordan-Wigner transformation~\cite{Lieb1962}
\begin{eqnarray}\label{JWT}
\sigma^+_j\equiv(\sigma^x_j+i\sigma^y_j)/2=c^\dag_j T_j,~\sigma^z_j=2c^\dag_j c_j-1,
\end{eqnarray}
where $c^\dag_j$ creates a spinless fermion on site $j$ and $T_j$ is the Jordan-Wigner string
\begin{eqnarray}\label{Tj}
T_j=e^{i\pi\sum^{j-1}_{l=1}c^\dag_l c_l}.
\end{eqnarray}
Note that $T_{N+1}=e^{i\pi\sum^{N}_{l=1}c^\dag_l c_l}$ is the fermion parity operator and it can be checked that $T_{N+1}$ is a conserved quantity with eigenvalues $\pm1$.
The Jordan-Wigner transformation maps the spin model $H_{\mathrm{QIM}}$ into a spinless fermion model
\begin{eqnarray}\label{HF}
H_{\mathrm{F}}&=&-\sum^{N-1}_{j=1}(c^\dag_j c_{j+1}+c^\dag_{j+1}c_j+c^\dag_j c^\dag_{j+1}+c_{j+1}c_j)\nonumber\\
&&+(c^\dag_Nc_1+c^\dag_1c_N+c^\dag_Nc^\dag_1+c_1c_N)T_{N+1}\nonumber\\
&&-2g\sum^N_{j=1}c^\dag_jc_j+gN,
\end{eqnarray}
where we have separated out the bulk and boundary contributions of the hopping and pairing terms.
\par Since $T_{N+1}$ is conserved and squares to 1, we can separately diagonalize $H_{\mathrm{F}}$ in the two subspaces with even ($T_{N+1}=1$) and odd ($T_{N+1}=-1$) fermion parity. In turn, we define two projection operators $P_+=(1+ T_{N+1})/2$ and $P_-=(1- T_{N+1})/2$, which project onto subspaces with even and odd number of fermions, respectively. Using the relation $P_{+}+P_-=1$, the fermion annihilation operator $c_j$ can be written as
\begin{eqnarray}\label{cj}
c_j&=&(P_++P_-)c_j(P_++P_-)\nonumber\\
&=&P_+c_jP_-+P_-c_jP_+.
\end{eqnarray}
Typical quadratic terms can be expressed as
\begin{eqnarray}\label{cca}
c^\dag_ic_j&=&P_+c^\dag_ic_jP_++P_-c^\dag_ic_jP_-,\nonumber\\
c^\dag_ic^\dag_j&=&P_+c^\dag_ic^\dag_jP_++P_-c^\dag_ic^\dag_jP_-.
\end{eqnarray}
Using the above equations and note that $T_{N+1}=P_+-P_-$ and $P_+P_-=P_-P_+=0$, the fermionic Hamiltonian $H_{\mathrm{F}}$ can be rewritten as
\begin{eqnarray}\label{HFPP}
H_{\mathrm{F}}&=& \sum_{\sigma=\pm}P_\sigma H_{\mathrm{F},\sigma}P_\sigma,\\
H_{\mathrm{F},\sigma}&=&- \sum^{N-1}_{j=1} (c^\dag_j c_{j+1}+c^\dag_j c^\dag_{j+1}+\mathrm{H.c.}) \nonumber\\
&&+\sigma  (c^\dag_Nc_1+c^\dag_Nc^\dag_1+\mathrm{H.c.})  \nonumber\\
&&-2g\sum^N_{j=1}  c^\dag_jc_j + gN,
\end{eqnarray}
where $\mathrm{H.c.}$ denotes the Hermitian conjugate. We can diagonalize $H_{\mathrm{F},+}$ and $H_{\mathrm{F},-}$ separately since the two subspaces are orthogonal to each other. To proceed, we focus on the $\sigma$-subspace and define
\begin{eqnarray}\label{cN1}
c_{N+1}\equiv -\sigma c_1,
\end{eqnarray}
then the boundary terms in $H_{\mathrm{F},\sigma}$ can be absorbed into the bulk ones to form a compact expression
\begin{eqnarray}\label{Hcj}
H_{\mathrm{F},\sigma}&=&- \sum^{N}_{j=1}\left[ (c^\dag_j c_{j+1}+c^\dag_j c^\dag_{j+1}+\mathrm{H.c.})+2gc^\dag_jc_j\right] + gN.\nonumber\\
\end{eqnarray}
Note that the spatial index $j$ in the sum now runs from $1$ to $N$, which suggests us to introduce the following Fourier transforms
\begin{eqnarray}\label{cjFT}
c_{j}&=& \frac{e^{i\pi/4}}{\sqrt{N}}\sum_{k\in K_\sigma}e^{ikj}c_{k\sigma},\nonumber\\
c_{k\sigma}&=& \frac{e^{-i\pi/4}}{\sqrt{N}}\sum^N_{j=1}e^{-ikj}c_{j},
\end{eqnarray}
where $\{c_{k\sigma}\}$ are the fermion annihilation operators in the Fourier space and a factor of $e^{i\pi/4}$ is introduced for later convenience. From Eq.~(\ref{cN1}), the allowed wave numbers for $\sigma=+$ and $-$ survive in the sets
\begin{eqnarray}\label{K+}
K_+&=&\left\{-\pi+\frac{\pi}{N},-\pi+\frac{3\pi}{N},\cdots,-\frac{\pi}{N},\frac{\pi}{N},\cdots,\pi-\frac{\pi}{N}\right\},\nonumber\\
\end{eqnarray}
and
\begin{eqnarray}\label{K-}
K_-&=&\left\{-\pi,-\pi+\frac{2\pi}{N},\cdots,0,\cdots,\pi-\frac{2\pi}{N}\right\},
\end{eqnarray}
respectively. Note that $K_+$ is symmetric with respect to $k\to -k$ and only $K_-$ includes the two special modes $k=-\pi$ and $0$ satisfying $k=-k$. It is easy to check from Eqs.~(\ref{cjFT}), (\ref{K+}) and (\ref{K-}) that the Fourier modes in the $\sigma$-subspace satisfy the usual anticommutation relations
\begin{eqnarray}
\{c_{k\sigma},c_{k'\sigma}\}=0,~~\{c_{k\sigma},c^\dag_{k'\sigma}\}=\delta_{kk'}.
\end{eqnarray}
However, for two wave numbers $k$ and $k'$ from two distinct sets, we have from Eq.~(\ref{cjFT}) (define $\bar{\sigma}=-\sigma$)
\begin{eqnarray}\label{ckkp}
\{c_{k\sigma},c^\dag_{k'\bar{\sigma}}\}=\frac{2}{N}\frac{ 1}{e^{i(k-k')}-1}.
\end{eqnarray}
Such kind of commutation relations can be safely avoided provided we concentrate on each subspace separately. However, as we will see in Sec.~\ref{IV}, they will play an important role in the calculation of the time evolution of the longitudinal magnetization. From Eq.~(\ref{ckkp}), the vacuum expectation value of $c_{k\sigma} c^\dag_{k'\bar{\sigma}}$ is simply
\begin{eqnarray}\label{fkkp}
f_{kk'}\equiv\langle\mathrm{vac}|c_{k\sigma} c^\dag_{k'\bar{\sigma}}|\mathrm{vac}\rangle=\frac{2}{N}\frac{ 1}{e^{i(k-k')}-1},
\end{eqnarray}
where $|\mathrm{vac}\rangle$ is the common vacuum of $c_j$ for all $j$.
\par We now insert Eq.~(\ref{cjFT}) into Eq.~(\ref{Hcj}) and note that $\frac{1}{N}\sum^N_{j=1}e^{i(k-k')j}=\delta_{k,k'}$ holds for $k,k'\in K_\sigma$, then straightforward calculation gives the Bogoliubov-de Gennes Hamiltonian
\begin{eqnarray}
H_{\mathrm{F},+}&=&\sum_{k>0,k\in K_+}H_{k+},\\
\label{HF+}
H_{\mathrm{F},-}&=&\sum_{k>0,k\in K_-}H_{k-}+\frac{1}{2}\left(H_{-\pi,-}+H_{0,-}\right),
\label{HF-}
\end{eqnarray}
where
\begin{eqnarray}
H_{k,\sigma}&\equiv& -2(c^\dag_{k\sigma},c_{-k,\sigma})\left(
                                                     \begin{array}{cc}
                                                       \cos k+g &\sin k \\
                                                       \sin k & -\cos k-g \\
                                                     \end{array}
                                                   \right)\nonumber\\
                                                   &&\left(
                                                       \begin{array}{c}
                                                         c_{k\sigma} \\
                                                         c^\dag_{-k,\sigma} \\
                                                       \end{array}
                                                     \right).
\end{eqnarray}
Note that $H_{k,-}$ is also well-defined for $k=-\pi$ and $0$, for which we have
\begin{eqnarray}\label{Hpi0-}
H_{-\pi,-}&=&2(1-g)(2c^\dag_{-\pi,-}c_{-\pi,-}-1),
\end{eqnarray}
and
\begin{eqnarray}\label{H0-}
H_{0,-}&=& -2(1+g)(2c^\dag_{0,-}c_{0,-}-1).
\end{eqnarray}
The diagonalization of $H_{\mathrm{QIM}}$ is now equivalent to diagonalizing each of the mode Hamiltonians, $\{H_{k,\sigma}\}$. Since the two special modes $k=-\pi$ and $0$ only appear in $K_-$, we can write $H_{-\pi/0}=H_{-\pi/0,-}$ and $c_{-\pi/0}=c_{-\pi/0,-}$, etc., without causing confusions.
\subsection{Normal modes: $k\neq -\pi$ and $k\neq 0$}
\par For $k\neq -\pi$ and $k\neq 0$, there always exists an \emph{opposite} and \emph{distinct} element $-k$ for each $k>0$ in both $K_+$ and $K_-$. The even subspace for mode $k$ (do not be confused with the global even subspace defined by $\sigma=+1$) is spanned by $\{|\mathrm{vac}\rangle_{k,\sigma},|k,-k\rangle_{\sigma}=c^\dag_{k,\sigma}c^\dag_{-k,\sigma}|\mathrm{vac}\rangle_{k,\sigma}\}$, where $|\mathrm{vac}\rangle_{k,\sigma}$ is the vacuum state of $c_{ k,\sigma}$ and $c_{- k,\sigma}$. It is easy to check that in this two-dimensional subspace $H_{k,\sigma}$ has the matrix form
\begin{eqnarray}\label{mHe}
\mathcal{H}^{(\mathrm{e})}_{k,\sigma}&=&2 \left(
                       \begin{array}{cc}
                         \cos k+ g & - \sin k \\
                         - \sin k & - \cos k- g \\
                       \end{array}
                     \right).
\end{eqnarray}
Similarly, the odd subspace for mode $k$ is spanned by $\{|-k\rangle_{\sigma}=c^\dag_{-k,\sigma}|\mathrm{vac}\rangle_{k,\sigma},|k\rangle_\sigma=c^\dag_{k,\sigma}|\mathrm{vac}\rangle_{k,\sigma}\}$, it turns out that for $k\neq -\pi$ and $k\neq0$ we always have
\begin{eqnarray}
\mathcal{H}^{(\mathrm{o})}_{k,\sigma}&=&  \left(
                       \begin{array}{cc}
                         0 & 0 \\
                         0 & 0 \\
                       \end{array}
                     \right).
\end{eqnarray}
In other words, $|-k\rangle_{\sigma}$ and $|k\rangle_{\sigma}$ are two degenerate eigenstates of $H_{k,\sigma}$ with zero eigenenergy.
\par The ground and excited states of $\mathcal{H}^{(\mathrm{e})}_{k,\sigma}$ are
\begin{eqnarray}
|G^{(\mathrm{gs})}_k\rangle_\sigma&=&\cos\frac{\theta_k}{2}|0\rangle_{k,\sigma}+\sin\frac{\theta_k}{2}|k,-k\rangle_{k,\sigma},\\
\label{GSe}
|G^{(\mathrm{ex})}_k\rangle_\sigma&=&\sin\frac{\theta_k}{2}|0\rangle_{k,\sigma}-\cos\frac{\theta_k}{2}|k,-k\rangle_{k,\sigma},
\label{EXe}
\end{eqnarray}
with the corresponding eigenenergies given by
\begin{eqnarray}
E^{(\mathrm{gs})}_{k\sigma}&=&-\Lambda_k,~~E^{(\mathrm{ex})}_{k\sigma}= \Lambda_k,\nonumber\\
\Lambda_k&=&2\sqrt{g^2+2g\cos k+1},
\label{Lambdak}
\end{eqnarray}
where
\begin{eqnarray}\label{CS2}
\sin\frac{\theta_k}{2}& =&\frac{2\sin k}{\sqrt{(2 \sin k)^2+(\Lambda_k-2\cos k-2g )^2}},\nonumber\\
\cos\frac{\theta_k}{2}& =&\frac{ \Lambda_k-2\cos k-2g}{\sqrt{(2 \sin k)^2+(\Lambda_k-2\cos k-2g )^2}}.
\end{eqnarray}
\par It is useful to note that $\Lambda_k> 0$ for all the normal modes and the lowest two excitation energies (i.e. $\Lambda_k$) are achieved for
\begin{eqnarray}
k^{(1)}_{\mathrm{e}}=\pi-\frac{\pi}{N},~k^{(2)}_{\mathrm{e}}=\pi-\frac{3\pi}{N},~(\sigma=+1),\\
k^{(1)}_{\mathrm{o}}=\pi-\frac{2\pi}{N},~k^{(2)}_{\mathrm{o}}=\pi-\frac{4\pi}{N},~(\sigma=-1).
\end{eqnarray}
We see that the ground state for a normal mode $k$ is just $|G^{(\mathrm{gs})}_k\rangle_\sigma$, which is an \emph{even} state. We thus conclude that \emph{the ground state is even for any normal mode $k$}.
\par In summary, the mode Hamiltonian $H_{k,\sigma}$ has four eigenvalues (in descending order)
\begin{eqnarray}
 \Lambda_k,~0,~0,~-\Lambda_k,\nonumber
\end{eqnarray}
and the corresponding eigenvectors are
\begin{eqnarray}
|G^{(\mathrm{ex})}_k\rangle_\sigma,~|k\rangle_\sigma,~|-k\rangle_\sigma,~|G^{(\mathrm{gs})}_k\rangle_\sigma,\nonumber
\end{eqnarray}
with fermion parities
\begin{eqnarray}
\rm{even,~odd,~odd,~even}.\nonumber
\end{eqnarray}
\par It is worth mentioning that for the classical Ising model with $g=0$, the dispersion becomes a constant $\Lambda_k=2$. From Eq.~(\ref{CS2}) we have
\begin{eqnarray}
\sin\frac{\theta_k|_{g=0}}{2}& =&\frac{ \sin k}{\sqrt{(  \sin k)^2+(1- \cos k  )^2}},\nonumber\\
\cos\frac{\theta_k|_{g=0}}{2}& =&\frac{ 1- \cos k }{\sqrt{( \sin k)^2+(1- \cos k )^2}}.\nonumber
\end{eqnarray}
Note that for normal modes satisfying $0<k<\pi$, we always have $\sin \frac{k}{2}>0$, so that $\sqrt{(  \sin k)^2+(1- \cos k  )^2}=\sqrt{2-2\cos k}=2 \sin\frac{k}{2} $, giving
\begin{eqnarray}\label{CS2g0}
\sin\frac{\theta_k|_{g=0}}{2}& =&\cos\frac{k}{2},\nonumber\\
\cos\frac{\theta_k|_{g=0}}{2}& =&\sin\frac{k}{2}.
\end{eqnarray}
\subsection{Special modes in the odd subspace}
\par Special attention must be paid to $k=-\pi$ and $0$, for which neither $|k,-k\rangle$ (it vanishes) nor $\{|-k\rangle,|k\rangle\}$ (the two are the same) is well defined. In fact, from Eqs.~(\ref{Hpi0-}) and (\ref{H0-}) we see that $H_{-\pi}$ and $H_{0}$ can be written in the basis $\{|\mathrm{vac}\rangle_{-\pi},|-\pi\rangle=c^\dag_{-\pi}|\mathrm{vac}\rangle_{-\pi}\}$ and $\{|\mathrm{vac}\rangle_{0},|0\rangle=c^\dag_{0}|\mathrm{vac}\rangle_{0}\}$ as
\begin{eqnarray}
\mathcal{H}_{-\pi}&=& \left(
                \begin{array}{cc}
                  -2(1-g) & 0 \\
                  0 & 2(1-g) \\
                \end{array}
              \right),
\end{eqnarray}
and
\begin{eqnarray}
\mathcal{H}_{0}&=&\left(
                \begin{array}{cc}
                   2(1+g) & 0 \\
                  0 & -2(1+g) \\
                \end{array}
              \right),
\end{eqnarray}
which are already in the diagonal form.
\section{Ground state}\label{III}
\par It is known that the ground state of $H_{\mathrm{QIM}}$ with even number of sites has an \emph{even} fermion parity for $g> 0$~\cite{Dzig2005,Sire2010,Rams2014,Nishimori2015,Katsura2012}. This can be demonstrated either in the (real-space) spin representation of the model with the help of the Perron-Frobenius theorem~\cite{Sire2010,Katsura2012}, or through a momentum-space analysis by comparing the eigenenergies of the even and odd sub-ground states in the two parity sectors~\cite{Rams2014,Nishimori2015}. Here, an even (odd) sub-ground state (SGS) means the lowest-energy fermionic state of the fermion Hamiltonian $H_{\mathrm{F},+}$ ($H_{\mathrm{F},-})$ having even (odd) number of fermions. In Appendix~\ref{app1}, we review previous results of determining the ground-state fermion parity by using the Perron-Frobenius theorem, which actually provides a more straightforward argument than the momentum-space analysis. Nevertheless, for our purpose, let us first identity the two SGSs in the two parity sectors in a rigorous way.
\subsection{Sub-ground states in the two parity sectors}
\par Since all the modes in $K_+$ are normal modes, the even SGS is simply
\begin{eqnarray}\label{GSeven}
|G_+\rangle=\prod_{k>0,k\in K_+}|G^{(\mathrm{gs})}_k\rangle_+.
\end{eqnarray}
The corresponding SGS energy is
\begin{eqnarray}
E_{G,+}=-\sum_{k>0,k\in K_+}\Lambda_k.
\end{eqnarray}
\par Note that any physical state in the even subspace must have an even fermion parity, so the first excited state in this subspace can be obtained by either exciting a single mode $q$ from its ground state $|G^{(\mathrm{gs})}_q\rangle_+$ to its (even) excited state $|G^{(\mathrm{ex})}_q\rangle_+$ (with eigenenergy $+\Lambda_q$), or by exciting two modes $q_1$ and $q_2$ to their (odd) excited states $|\pm q_1\rangle_+$ and $|\pm q_2\rangle_+$ (with zero eigenenergy). The minimal energy cost for the former case is $2\Lambda_{k^{(1)}_{\mathrm{e}}}$, while for the latter case is $\Lambda_{k^{(1)}_{\mathrm{e}}}+\Lambda_{k^{(2)}_{\mathrm{e}}}>2\Lambda_{k^{(1)}_{\mathrm{e}}}$. Therefore, the first excited state in the even subspace is
\begin{eqnarray}
|\mathrm{ex}_+\rangle=|G^{(\mathrm{ex})}_{k^{(1)}_{\mathrm{e}}}\rangle_+ \prod_{k>0,k\in K_+,k\neq k^{(1)}_{\mathrm{e}} }|G^{(\mathrm{gs})}_k\rangle_+,
\end{eqnarray}
with eigenenergy
\begin{eqnarray}\label{1stex}
E_{\mathrm{ex},+}=E_{G,+}+2\Lambda_{k^{(1)}_{\mathrm{e}}}.
\end{eqnarray}
Since $\Lambda_{k^{(1)}_{\mathrm{e}}}$ is always positive for finite $N$, equation~(\ref{1stex}) tells us that \emph{the even SGS is nondegenerate in the even subspace}.
\par The situation in the odd subspace is a little subtle due to the appearance of the special modes $k=-\pi$ and $0$. We define $|\phi_{\mathrm{e}}\rangle$ ($|\phi_{\mathrm{o}}\rangle$) to be the lowest-energy state with even (odd) fermion parity and \emph{made up of the normal modes} in the odd sector. It is apparent that
\begin{eqnarray}
|\phi_{\mathrm{e}}\rangle &=&\prod_{k>0,k\in K_-}|G^{\mathrm{gs}}_k\rangle_-,\nonumber\\
|\phi_{\mathrm{o}}\rangle &=&|\pm k^{(1)}_{\mathrm{o}}\rangle_- \prod_{k>0,k\in K_-, k\neq k^{(1)}_{\mathrm{o}}}|G^{\mathrm{gs}}_k\rangle_-,~
\end{eqnarray}
which possess energies
\begin{eqnarray}
\mathcal{E}_{\mathrm{e}}&=&-\sum_{k>0,k\in K_-}\Lambda_k,\nonumber\\
\mathcal{E}_{\mathrm{o}}&=&\mathcal{E}_{\mathrm{e}}+\Lambda_{k^{(1)}_{\mathrm{o}}}.
\end{eqnarray}
Given $|\phi_{\mathrm{e}}\rangle$ and $|\phi_{\mathrm{o}}\rangle$, there are four possible ways to construct a physical eigenstate in the odd subspace:
\begin{eqnarray}
|\psi_1\rangle&=&|\mathrm{vac}\rangle_{-\pi}|\mathrm{vac}\rangle_{0}|\phi_{\mathrm{o}}\rangle,\nonumber\\
|\psi_2\rangle&=&|\mathrm{vac}\rangle_{-\pi}|0\rangle |\phi_{\mathrm{e}}\rangle,\nonumber\\
|\psi_3\rangle&=&|-\pi\rangle |0\rangle |\phi_{\mathrm{o}}\rangle,\nonumber\\
|\psi_4\rangle&=&|-\pi\rangle |\mathrm{vac}\rangle_{0} |\phi_{\mathrm{e}}\rangle,\nonumber
\end{eqnarray}
with the corresponding eigenenergies
\begin{eqnarray}
E_1&=&2g+\mathcal{E}_{\mathrm{o}},\nonumber\\
E_2&=&-2+\mathcal{E}_{\mathrm{e}},\nonumber\\
E_3&=&-2g+\mathcal{E}_{\mathrm{o}},\nonumber\\
E_4&=&2+\mathcal{E}_{\mathrm{e}}.\nonumber
\end{eqnarray}
Since $g\geq 0$, the odd SGS should be either $|\psi_2\rangle$ or $|\psi_3\rangle$. The energy difference between these two states is
\begin{eqnarray}
E_2-E_3&=&2(g-1)-2\sqrt{g^2-2g\cos\frac{2\pi}{N}+1}.\nonumber
\end{eqnarray}
For $0\leq g\leq 1$, it is obvious that $E_2-E_3<0$; for $g>1$, $E_2-E_3$ is non-positive as $0\leq\cos\frac{2\pi}{N}<1$ for $N\geq 4$. Thus, the odd SGS must be $|\psi_2\rangle$ for all $g\geq0$, i.e.,
\begin{eqnarray}\label{GSodd}
|G_-\rangle=|\mathrm{vac}\rangle_{-\pi}|0\rangle\prod_{k>0,k\in K_-}|G^{(\mathrm{gs})}_k\rangle_-.
\end{eqnarray}
The corresponding SGS energy is
\begin{eqnarray}
E_{G,-}&=& -\sum_{k>0,k\in K_-}\Lambda_k-2.
\end{eqnarray}
\subsection{The global ground state}
\par The true ground state of $H_{\mathrm{QIM}}$, $|\psi_G\rangle$, must be either $|G_+\rangle$ or $|G_-\rangle$. We already know that $H_{\mathrm{QIM}}$ has a unique ground state with even fermion parity for even $N$ (see Appendix~\ref{app1}). In addition, we have shown that the even SGS $|G_+\rangle$ is gapped, so we must have
\begin{eqnarray}\label{PsiGG}
|\psi_{\mathrm{G}}\rangle=|G_+\rangle.
\end{eqnarray}
\par It is an interesting fact that the determination of the true ground state by directly comparing the two SGS energies, $E_{G,+}$ and $E_{G,-}$, is not obvious, especially in the ferromagnetic phase with $g<1$. We define the energy difference
\begin{eqnarray}\label{Dg}
\Delta(g)&\equiv&\frac{1}{2}( E_{G,-}-E_{G,+})\nonumber\\
&=&\sum^{N/2}_{j=1}\sqrt{g^2+2g\cos(2j-1)\alpha+1}\nonumber\\
&&-\sum^{N/2-1}_{j=1}\sqrt{g^2+2g\cos 2j\alpha+1}-1,
\end{eqnarray}
where $\alpha\equiv\pi/N$. Thus, the ground state will have even or odd fermion parity if $\Delta(g)>0$ or $\Delta(g)<0$, and Eq.~(\ref{PsiGG}) implies that the former is the case, i.e.,
\begin{eqnarray}\label{Dggl0}
\Delta(g)> 0,~~(g>0).
\end{eqnarray}
\begin{figure}
\includegraphics[width=.48\textwidth]{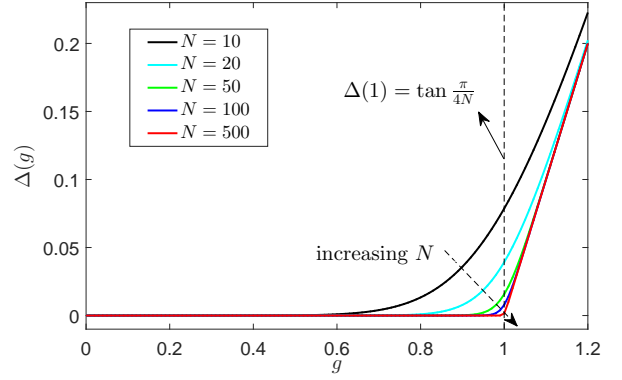}
\caption{The energy difference $\Delta (g)$ as a function of $g$ [see Eq.~(\ref{Dg})] for various values of $N$.}
\label{Fig1}
\end{figure}
It is obvious that $\Delta(0)=0$, meaning that the two SGSs $|G_+\rangle_{g=0}$ and $|G_- \rangle_{g=0}$ are degenerate in the absence of the magnetic field. We can also see this in the spin representation, where $H_{\mathrm{QIM}}$ reduces to the classical Ising model that has two degenerate ferromagnetic states aligned along the $\pm x$ directions. Another special point is $g=1$, for which we have $\Delta(1)=\tan\frac{\pi}{4N}$. In Fig.~\ref{Fig1} we plot the function $\Delta(g)$ for various values of $N$. We would like to mention that in Refs.~\cite{Rams2014,Nishimori2015} a closed-form expression for the energy difference $\Delta(g)$ is provided in the form of an integration.
\par
\begin{figure}
\hspace*{1.1cm}
\includegraphics[width=.70\textwidth]{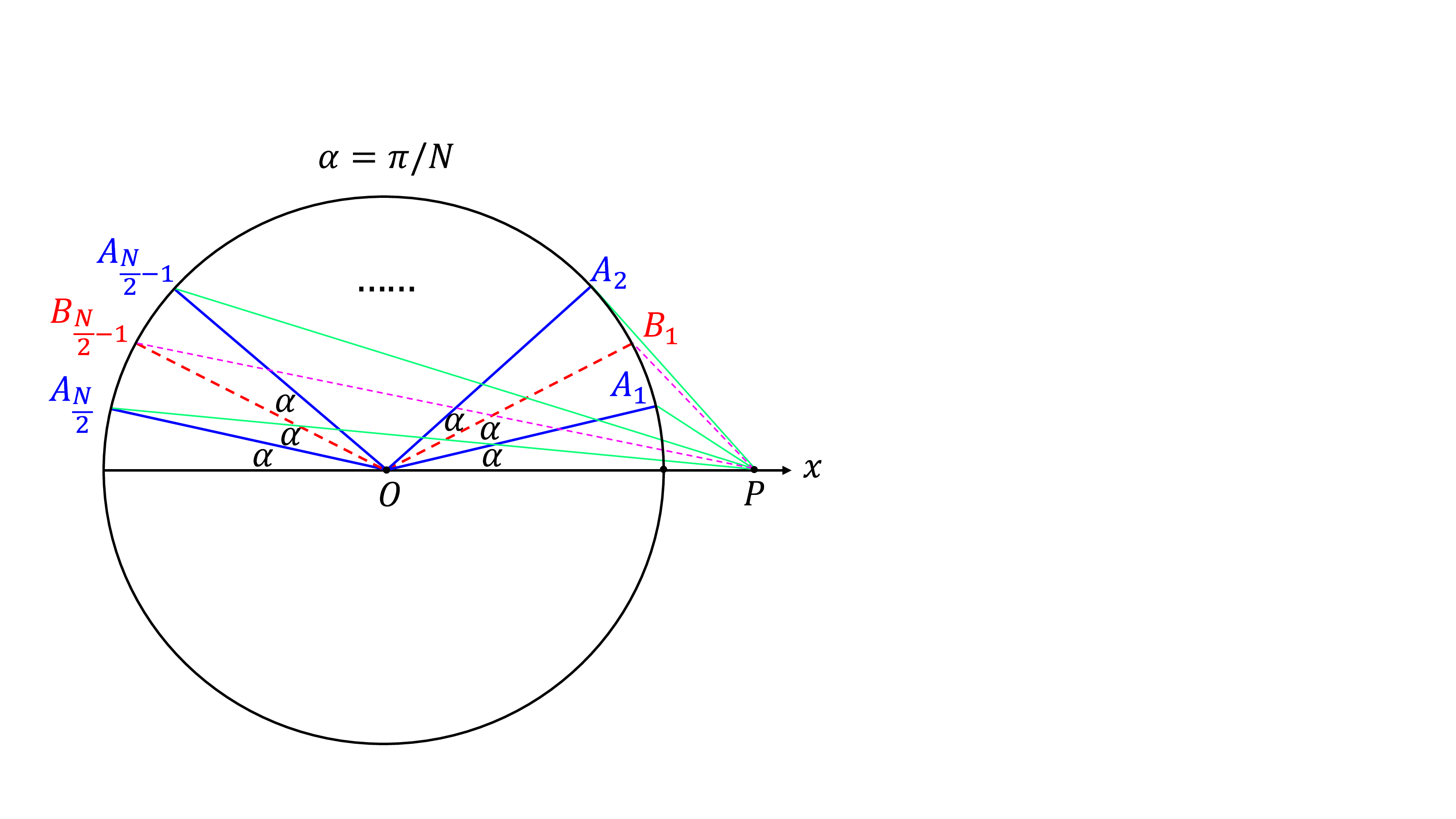}
\caption{The upper unit semicircle is divided into $N$ equal sectors, so that $\angle POA_j=(2j-1)\alpha$ ($j=1,2,\cdots,N/2$) and $\angle POB_j=2j\alpha$ ($j=1,2,\cdots,N/2-1$), where $\alpha=\pi/N$.}
\label{Fig2}
\end{figure}

 The direct proof of the inequality given by (\ref{Dggl0}) is itself an elementary mathematical problem having a geometric meaning. Let $P$ be a point on the positive $x$-axis and set $OP=x$. The upper half of the unit circle is divided into $N$ ($N$ is even) equal sectors with angle $\alpha=\pi/N$, such that $\angle POA_j=(2j-1)\alpha$ ($j=1,2,\cdots,N/2$) and $\angle POB_j=2j\alpha$ ($j=1,2,\cdots,N/2-1$) (see Fig.~\ref{Fig2}). Connecting $P$ to each point $A_j$ and $B_j$ defines the following quantity
\begin{eqnarray}\label{PAPB}
\Delta l(P)\equiv \sum^{N/2}_{j=1}PA_j-\sum^{N/2-1}_{j=1}PB_j,
\end{eqnarray}
which can be written as (using the law of cosine)
\begin{eqnarray}\label{PAPB1}
\Delta l(P)&=&\sum^{N/2}_{j=1}f_{2j-1}(x) -\sum^{N/2-1}_{j=1}f_{2j}(x),\nonumber\\
f_j(x)&\equiv& \sqrt{x^2-2x\cos (j\alpha)+1}.
\end{eqnarray}
It is easy to get the following relation between the energy difference $\Delta(x)$ and the quantity $\Delta l(P)$,
\begin{eqnarray}
\Delta(x)=\Delta l (P)-1,
\end{eqnarray}
so that the inequality (\ref{Dggl0}) is equivalent to
\begin{eqnarray}\label{DeltalP}
\Delta l (P)>1,~(OP>0)
\end{eqnarray}
We can straightforwardly prove inequality (\ref{DeltalP}) for $OP=x\geq 1$ (see Appendix~\ref{app2}). However, the situation is intricate in the region $0<x<1$ (i.e., $P$ inside the unit circle) where $\Delta l(P)-1$ becomes exponentially small for large $N$. Though there is numerical evidence that $\Delta l(P)-1>0$ also holds for $0<x<1$ (Fig.~\ref{Fig2}), a direct proof of this is not obvious since $f_j(x)$ is generally not a monotonic function of $x$ on $x\in[0,\cos\alpha)$.
\par Since we already showed that $\Delta(x)>0$ must hold for $x>0$, we thus proved the inequality given by (\ref{DeltalP}) for $0<x<1$ by combining the Perron-Frobenius theorem in matrix theory with the fermion representation of $H_{\mathrm{QIM}}$, which can be viewed as a kind of ``physical mathematics" in some sense.
\subsection{$g=0$: Separable ground states}
\par For $g=0$, the quantum Ising model $H_{\mathrm{QIM}}$ reduces to the classical Ising model $H_{\mathrm{CIM}}=-\sum^N_{j=1}\sigma^x_j\sigma^x_{j+1}$, which possesses two degenerate ground states
\begin{eqnarray}
|\psi^{\mathrm{(r)}}_{g=0}\rangle&=&|\rightarrow\rangle_1\cdots|\rightarrow\rangle_N,\label{FMright}\\
|\psi^{\mathrm{(l)}}_{g=0}\rangle&=&|\leftarrow\rangle_1\cdots|\leftarrow\rangle_N,\label{FMleft}
\end{eqnarray}
where $|\rightarrow\rangle_j$ ($|\leftarrow\rangle_j$) denotes the state with the $j$th spin pointing along the $+x$ ($-x$) direction. Actually, the classical Ising Hamiltonian given by $H_{\mathrm{CIM}}$ is a special factorization point of the XYZ model~\cite{Krumann,Muller1985,Cerezo2015,Katsura2015} at which the two spatially factorized ground states $|\psi^{(\mathrm{r/l})}_{g=0}\rangle$ are \emph{orthogonal} to each other (see Appendix~\ref{app3} for details).
\par Let us first look at the two ferromagnetic states in the real-space \emph{fermion representation}. Using the relation $\sigma^+_{j_1}\cdots \sigma^+_{j_n}|\downarrow\cdots\downarrow\rangle=c^\dag_{j_1}\cdots c^\dag_{j_n}|\mathrm{vac}\rangle$ which is valid for $j_1<\cdots<j_n$~\cite{PRA2014}, we can write $|\psi^{\mathrm{(r/l)}}_{g=0}\rangle$ in terms of the Jordan-Wigner fermions as
\begin{eqnarray}\label{FMrl}
|\psi^{\mathrm{(r/l)}}_{g=0}\rangle&=&\left(\frac{1}{\sqrt{2}}\right)^N\prod^N_{j=1}\left(1\pm c^\dag_j\right)|\mathrm{vac}\rangle.
\end{eqnarray}
However, neither $|\psi^{\mathrm{(r )}}_{g=0}\rangle$ nor $|\psi^{\mathrm{( l)}}_{g=0}\rangle$ has a definite fermion parity, as can be seen from the expansion
\begin{eqnarray}\label{FMrlexp}
&&|\psi^{\mathrm{(r/l)}}_{g=0}\rangle=\left(\frac{1}{\sqrt{2}}\right)^N\sum^{N/2}_{n=0}\sum_{j_1<\cdots<j_{2n}}c^\dag_{j_1}\cdots c^\dag_{j_{2n}}|\mathrm{vac}\rangle\nonumber\\
&&\pm\left(\frac{1}{\sqrt{2}}\right)^N\sum^{N/2}_{n=1}\sum_{j_1<\cdots<j_{2n-1}}c^\dag_{j_1}\cdots c^\dag_{j_{2n-1}}|\mathrm{vac}\rangle.
\end{eqnarray}
We now consider the following two cat states that are equally weighted superpositions of $|\psi^{\mathrm{(r )}}_{g=0}\rangle$ and $|\psi^{\mathrm{( l)}}_{g=0}\rangle$:
\begin{eqnarray}\label{cat+}
|\mathrm{cat}_{+}\rangle&=&\frac{1}{\sqrt{2}}(|\psi^{\mathrm{(r )}}_{g=0}\rangle+|\psi^{\mathrm{( l)}}_{g=0}\rangle)\nonumber\\
&=&\left(\frac{1}{\sqrt{2}}\right)^{N-1}\sum^{N/2}_{n=0}\sum_{j_1<\cdots<j_{2n}}c^\dag_{j_1}\cdots c^\dag_{j_{2n}}|\mathrm{vac}\rangle,\nonumber\\
\end{eqnarray}
and
\begin{eqnarray}\label{cat-}
|\mathrm{cat}_{-}\rangle&=&\frac{1}{\sqrt{2}}(|\psi^{\mathrm{(r )}}_{g=0}\rangle- |\psi^{\mathrm{( l)}}_{g=0}\rangle)\nonumber\\
&=&\left(\frac{1}{\sqrt{2}}\right)^{N-1}\sum^{N/2}_{n=1}\sum_{j_1<\cdots<j_{2n-1}}c^\dag_{j_1}\cdots c^\dag_{j_{2n-1}}|\mathrm{vac}\rangle.\nonumber\\
\end{eqnarray}
It is easy to see that $|\mathrm{cat}_{+}\rangle$ ($|\mathrm{cat}_{-}\rangle$) is a simultaneous eigenstate of $T_{N+1}$ and $H_{\mathrm{CIM}}$, with eigenvalues $+1$ ($-1$) and $-N$, respectively. The two expressions given by (\ref{cat+}) and (\ref{cat-}) seem at first sight complicated due to the appearance of a large number of fermion occupation states having a variety of different fermion numbers. As we will show below, the two states $|\mathrm{cat}_{\pm}\rangle$ actually admit simple forms in the momentum-space of the Jordan-Wigner fermions.
\par From Eqs.~(\ref{CS2g0}), (\ref{GSeven}), and (\ref{GSodd}), the degenerate even and odd SGSs of $H_{\mathrm{CIM}}$ read
\begin{eqnarray}
|G_+ \rangle_{g=0}&=&\prod_{k>0,k\in K_+}B_{k,+}|\mathrm{vac}\rangle_{k,+},\\
|G_- \rangle_{g=0}&=&|\mathrm{vac}\rangle_{-\pi}|0\rangle\prod_{k>0,k\in K_-}B_{k,-}|\mathrm{vac}\rangle_{k,-},
\end{eqnarray}
where we defined the (even) operator
\begin{eqnarray}
B_{k,\sigma}\equiv \sin\frac{k}{2}+\cos\frac{k}{2}c^\dag_{k,\sigma}c^\dag_{-k,\sigma}.
\end{eqnarray}
\par It is obvious that there is a finite energy gap of value 4 between the first excited state and the ground-state manifold of $H_{\mathrm{CIM}}$. By noting that $|\mathrm{cat}_{\sigma}\rangle$ and $|G_\sigma \rangle_{g=0}$ have identical fermion parity, $(-1)^\sigma$, the real-space state $|\mathrm{cat}_{\sigma}\rangle$ is expected to be equivalent to the momentum-space state  $|G_{\sigma}\rangle_{g=0}$, i.e.,
\begin{eqnarray}\label{catgs}
|\mathrm{cat}_{\sigma}\rangle&=&e^{i\theta_{\sigma}}|G_{\sigma}\rangle_{g=0},
\end{eqnarray}
where $e^{i\theta_\sigma}$ is a proportional constant that needs to be determined. To determine $e^{i\theta_+}$, we compare the coefficients of the vacuum state on both sides of $|\mathrm{cat}_{+}\rangle=e^{i\theta_{+}}|G_{+}\rangle_{g=0}$, yielding
\begin{eqnarray}
\left(\frac{1}{\sqrt{2}}\right)^{N-1}=e^{i\theta_+}\prod_{k>0,k\in K_+}\sin\frac{k}{2}.\nonumber
\end{eqnarray}
Using the identity $\omega^N+1=\prod^N_{j=1}\left(\omega-e^{i\pi\frac{2j-1}{N}}\right)$, it is easy to show that
\begin{eqnarray}
&&\prod_{k>0,k\in K_+}\sin\frac{k}{2}=\prod^{N/2}_{j=1}\sin\frac{(2j-1)\pi}{2N} \nonumber\\
&=&\sqrt{\frac{1}{2^N}\prod^{N/2}_{j=1}|1-e^{i\pi\frac{2j-1}{N}}|}=\left(\frac{1}{\sqrt{2}}\right)^{N-1},\nonumber
\end{eqnarray}
which gives $e^{i\theta_+}=1$, and hence
\begin{eqnarray}\label{catgs+}
|\mathrm{cat}_{+}\rangle&=&|G_{+}\rangle_{g=0}.
\end{eqnarray}
\par Similarly, by comparing the coefficients of the single-particle component of $|\mathrm{cat}_{-}\rangle=e^{i\theta_{-}}|G_{-}\rangle_{g=0}$, we have  $e^{i\theta_-}=e^{-i\pi/4}$, giving
\begin{eqnarray}\label{catgs-}
|\mathrm{cat}_{-}\rangle&=&e^{- i\pi/4} |G_{-}\rangle_{g=0}.
\end{eqnarray}
\par It is worth mentioning that the two relations given by Eqs.~(\ref{catgs+}) and (\ref{catgs-}) have been observed in Ref.~\cite{Dong} through numerical tests on small rings, and Eq.~(\ref{catgs-}) was explicitly proved in Ref.~\cite{Greiter2014} by transforming the momentum-space state $|G_-\rangle_{g=0}$ into the \emph{real space} and showing the consistency of the corresponding coefficients between $|\mathrm{cat}_{-}\rangle$ and the transformed  $|G_-\rangle_{g=0}$ with the help of mathematical induction. However, our proof of Eqs.~(\ref{catgs+}) and (\ref{catgs-}) is based on physical considerations and is more concise.
\section{Dynamics of the longitudinal magnetization under translationally invariant drivings}\label{IV}
\par After establishing the relationship between the two ferromagnetic states $|\psi^{\mathrm{(r/l)}}_{g=0}\rangle$ and the two momentum-space states $|G_\pm\rangle_{g=0}$, we are now ready to study the time evolution starting with one of the two ferromagnetic states, say $|\psi^{\mathrm{(r)}}_{g=0}\rangle$. In this section, we are interested in the calculation of the dynamics of the longitudinal magnetizations $M_x=\sum_i\sigma^x_i$ and $M_y=\sum_i\sigma^y_i$ when the system is prepared in $|\psi^{\mathrm{(r)}}_{g=0}\rangle$ and the driving Hamiltonian is translationally invariant, which guarantees that the time-evolved state is still expressible in the momentum space.
\subsection{General formalism}\label{Generalform}
\par From Eqs.~(\ref{cat+}), (\ref{cat-}), (\ref{catgs+}), and (\ref{catgs-}), we can write $|\psi^{\mathrm{(r)}}_{g=0}\rangle$ as
\begin{eqnarray}
|\psi^{\mathrm{(r)}}_{g=0}\rangle&=&\frac{1}{\sqrt{2}}\left(|G_+\rangle_{g=0}+e^{ -i\frac{\pi}{4}}|G_-\rangle_{g=0}\right)\nonumber\\
&=&\frac{1}{\sqrt{2}} \prod_{k>0,k\in K_+}B_{k,+}|\mathrm{vac}\rangle_{k+}\nonumber\\
&&+\frac{e^{- i\frac{\pi}{4}}}{\sqrt{2}}|\mathrm{vac}\rangle_{-\pi}|0\rangle \prod_{k>0,k\in K_-}B_{k,-}|\mathrm{vac}\rangle_{k-}.\nonumber\\
\end{eqnarray}
\par We assume that the (possibly time-dependent) driving Hamiltonian $H^{\mathrm{(driv)}}(t)$ is translationally invariant so that it can be written in the momentum space as
\begin{eqnarray}
H^{\mathrm{(driv)}}&=&\sum_{\sigma=\pm}P_\sigma H^{\mathrm{(driv)}}_{\sigma}P_\sigma,\nonumber\\
H^{\mathrm{(driv)}}_{+}&=&\sum_{k>0,k\in K_+}H^{\mathrm{(driv)}}_{k,+},\nonumber\\
\label{Hdiv+}
H^{\mathrm{(driv)}}_{-}&=&\sum_{k>0,k\in K_-}H^{\mathrm{(driv)}}_{k,-}+\frac{1}{2}\left(H^{\mathrm{(driv)}}_{-\pi}+H^{\mathrm{(driv)}}_{0}\right).\nonumber
\label{Hdiv-}
\end{eqnarray}
For most of cases of interest, the representation of $H^{\mathrm{(driv)}}_{-\pi}$ ($H^{\mathrm{(driv)}}_{0}$) in the basis $\{|\mathrm{vac}\rangle_{-\pi},|-\pi\rangle\}$ ($\{|\mathrm{vac}\rangle_{0},|0\rangle\}$) is diagonal:
\begin{eqnarray}
\mathcal{H}^{\mathrm{(driv)}}_{-\pi}&=&\left(
                                 \begin{array}{cc}
                                   h^{(-\pi)}_{1}(t) & 0 \\
                                  0 & h^{(-\pi)}_{2}(t) \\
                                 \end{array}
                               \right),\nonumber\\
      \mathcal{H}^{\mathrm{(driv)}}_{0}&=&\left(
                                 \begin{array}{cc}
                                   h^{(0)}_{1}(t) & 0 \\
                                  0 & h^{(0)}_{2}(t) \\
                                 \end{array}
                               \right).
\end{eqnarray}
The time-evolved state is therefore of the form
\begin{eqnarray}\label{psirt}
&&|\psi^{\mathrm{(r)}}(t)\rangle=\nonumber\\
&& \frac{1}{\sqrt{2}}\prod_{k>0,k\in K_+}\left[u^{(+)}_k(t)+v^{(+)}_k(t)c^\dag_{k,+}c^\dag_{-k,+}\right]|\mathrm{vac}\rangle_{k+}\nonumber\\
&&+\frac{e^{-i\frac{\pi}{4}}}{\sqrt{2}}e^{-\frac{i}{2}\int^t_0ds\left[h^{(-\pi)}_{1}(s)+h^{(0)}_{2}(s)\right]}|\mathrm{vac}\rangle_{-\pi}|0\rangle\nonumber\\
&&\prod_{k>0,k\in K_-} \left[u^{(-)}_k(t)+v^{(-)}_k(t)c^\dag_{k,-}c^\dag_{-k,-}\right]|\mathrm{vac}\rangle_{k-},
\end{eqnarray}
where the time-dependent coefficients $u^{(\sigma)}_k(t)$ and $v^{(\sigma)}_k(t)$ satisfy the Schr\"odinger equation
\begin{eqnarray}\label{modeeq}
i\left(
   \begin{array}{c}
     \dot{u}^{(\sigma)}_k \\
     \dot{v}^{(\sigma)}_k \\
   \end{array}
 \right)=\mathcal{H}^{\mathrm{(driv)}}_{k,\sigma}(t)\left(
   \begin{array}{c}
     u^{(\sigma)}_k \\
     v^{(\sigma)}_k \\
   \end{array}
 \right),
\end{eqnarray}
with $\mathcal{H}^{\mathrm{(driv)}}_{k,\sigma}(t)$ being the representation of $H^{\mathrm{(driv)}}_{k,\sigma}(t)$ in the basis $\{|\mathrm{vac}\rangle_{k,\sigma},c^\dag_{k,\sigma}c^\dag_{-k,\sigma}|\mathrm{vac}\rangle_{k,\sigma}\}$, and the initial conditions read
\begin{eqnarray}
\left(
   \begin{array}{c}
     u^{(\sigma)}_k(t=0) \\
     v^{(\sigma)}_k(t=0) \\
   \end{array}
 \right)=\left(
   \begin{array}{c}
     \sin\frac{k}{2} \\
     \cos\frac{k}{2} \\
   \end{array}
 \right).
\end{eqnarray}
\par We are now facing the problem of calculating the expectation value of the longitudinal magnetizations
\begin{eqnarray}
M_{x/y}(t)&=&\langle  \psi^{\mathrm{(r)}}(t) |\sum^N_{j=1}\sigma^{x/y}_j|\psi^{\mathrm{(r)}}(t)\rangle,
\end{eqnarray}
which seems challenging at first sight because the state $|\psi^{\mathrm{(r)}}(t)\rangle$ is a momentum-space state containing components with both fermion parities, while $\sum^N_{j=1}\sigma^{x/y}_j$ are real-space operators and, more importantly, they break the fermion parity.
\par Thanks to the translational invariance of the initial state $|\psi^{\mathrm{(r)}}_{g=0}\rangle$ and the driving Hamiltonian $H^{\mathrm{(driv)}}$, the expectation values of each $\sigma^{x/y}_j$ should be independent of $j$ and we need only to calculate, for example, the magnetizations of the first site, so that
\begin{eqnarray}\label{Mxy}
M_{x}(t)&=&N\langle  \psi^{\mathrm{(r)}}(t) | \sigma^x_1|\psi^{\mathrm{(r)}}(t)\rangle\nonumber\\
&=&N\langle  \psi^{\mathrm{(r)}}(t) | (c_1+c^\dag_1)|\psi^{\mathrm{(r)}}(t)\rangle,\nonumber\\
M_{y}(t)&=&N\langle  \psi^{\mathrm{(r)}}(t) | \sigma^y_1|\psi^{\mathrm{(r)}}(t)\rangle\nonumber\\
&=&N\langle  \psi^{\mathrm{(r)}}(t) | i(c_1-c^\dag_1)|\psi^{\mathrm{(r)}}(t)\rangle.
\end{eqnarray}
To proceed, let us write $c_1$ in the momentum space as
\begin{eqnarray}
c_1&=&\sum_{\sigma=\pm}P_{-\sigma}\frac{e^{i\pi/4}}{\sqrt{N}}\sum_{k\in K_\sigma}e^{ik}c_{k\sigma}P_\sigma,
\end{eqnarray}
where we used Eqs.~(\ref{cj}) and (\ref{cjFT}).
From the above equation and the explicit form of $|\psi^{\mathrm{(r)}}(t)\rangle$ given by Eq.~(\ref{psirt}), we obtain after a straightforward, but lengthy calculation
\begin{widetext}
\begin{eqnarray}\label{lengthy}
&&\langle\psi^{\mathrm{(r)}}(t)|c_1|\psi^{\mathrm{(r)}}(t)\rangle\nonumber\\
&=&\frac{1}{2\sqrt{N}}e^{-i\gamma(t)} \left(\prod_{p>0,p\in K_+}~_+\langle X_{p}|\right)\left(|\mathrm{vac}\rangle_{-\pi}\prod_{k>0,k\in K_-} |X_k\rangle_-\right)\nonumber\\
&&+  \frac{1}{2\sqrt{ N}}e^{-i\gamma(t)}\left( \prod_{p>0,p\in K_+}~_+\langle X_{p}|\right) \left(\sum_{k'>0,k'\in K_-}v^{(-)}_{k'}(e^{ik'}|-k'\rangle_--e^{-ik'}|k'\rangle_- )|\mathrm{vac}\rangle_{-\pi}|0\rangle\prod_{k>0,k\in K_-,k(\neq k')}|X_k\rangle_- \right) \nonumber\\
&&+\frac{i}{2\sqrt{N}}e^{ i\gamma(t)} \left(~_{-\pi}\langle\mathrm{vac}|\langle 0|\prod_{k>0,k\in K_-}~_-\langle X_k| \right)\left(\sum_{p'>0,p'\in K_+}v^{(+)}_{p'}(e^{ip'}|-p'\rangle_+-e^{-ip'}|p'\rangle_+)\prod_{p>0,p\in K_+,p(\neq p')}|X_p\rangle_+\right),\nonumber\\
\end{eqnarray}
\end{widetext}
where we have defined
\begin{eqnarray}
\gamma(t)&\equiv&\frac{1}{2}\int^t_0ds\left[h^{(-\pi)}_{1}(s)+h^{(0)}_{2}(s)\right],\nonumber\\
|X_k\rangle_{\sigma}&\equiv& \left[u^{(\sigma)}_k+v^{(\sigma)}_kc^\dag_{k,\sigma}c^\dag_{-k,\sigma}\right]|\mathrm{vac}\rangle_{k\sigma}.
\end{eqnarray}
\par Even though we can obtain the mode state $|X_k\rangle_{\sigma}$ by solving Eq.~(\ref{modeeq}), we are forced to evaluate the inner products between two product states within \emph{distinct} fermion parity sectors appearing in Eq.~(\ref{lengthy}). In general, we have to calculate
\begin{eqnarray}
I_{m,n}=\prod^m_{l=1, p_l>0, p_l\in K_\sigma}~_\sigma\langle X_{p_l}|\prod^n_{j=1, k_j>0, k_j\in K_{\bar{\sigma}}}|X_{k_j}\rangle_{\bar{\sigma}}.
\end{eqnarray}
We now use the following trick to write $|X_k\rangle_{\sigma}$ as
\begin{eqnarray}\label{Xk}
|X_k\rangle_{\sigma}&=& \frac{1}{v^{(\sigma)}_k}\xi^\dag_{k\sigma}\eta^\dag_{k\sigma}|\mathrm{vac}\rangle_{k,\sigma},
\end{eqnarray}
where
\begin{eqnarray}
\xi^\dag_{k\sigma}&\equiv& u^{(\sigma)}_kc_{k,\sigma}-v^{(\sigma)}_kc^\dag_{-k,\sigma},\nonumber\\
\eta^\dag_{k\sigma}&\equiv& u^{(\sigma)}_kc_{-k,\sigma}+v^{(\sigma)}_kc^\dag_{k,\sigma},
\end{eqnarray}
are two time-dependent effective Bogoliubov fermions. From the orthonormal condition $u^{(\sigma)*}_k(t)u^{(\sigma)}_{k'}(t) +v^{(\sigma)*}_k(t)v^{(\sigma)}_{k'}(t)=\delta_{kk'}$, it is easy to see that $\{\xi^\dag_{k\sigma}\}$ and $\{\eta^\dag_{k\sigma}\}$ satisfy the usual anticommutation relations of fermions
\begin{eqnarray}
\{\xi_{k\sigma},\xi^\dag_{k'\sigma}\}&=&\delta_{kk'},~\{\eta_{k\sigma},\eta^\dag_{k'\sigma}\}=\delta_{kk'},\nonumber\\
\{\xi^\dag_{k\sigma},\eta^\dag_{k'\sigma}\}&=&\{\xi_{k\sigma},\eta_{k'\sigma}\}=\{\xi_{k\sigma},\eta^\dag_{k'\sigma}\}=0.
\end{eqnarray}
\par Using Eq.~(\ref{Xk}), the inner product $I_{m,n}$ can be expressed as the vacuum expectation value of a product of $2(m+n)$ Bogoliubov fermions
\begin{eqnarray}
I_{m,n}&=&\frac{(-1)^m}{\prod^m_{l=1}v^{(\sigma)*}_{p_l}\prod^n_{j=1}v^{(\bar{\sigma})}_{k_j}} \langle\mathrm{vac}|\prod^m_{l=1,p_l>0,p_l\in K_\sigma}\xi_{p_l,\sigma}\eta_{p_l,\sigma}\nonumber\\
&&\prod^n_{j=1,k_j>0,k_j\in K_{\bar{\sigma}}}\xi^\dag_{k_j,\bar{\sigma}}\eta^\dag_{k_j,\bar{\sigma}}|\mathrm{vac}\rangle.
\end{eqnarray}
Since all the fermion operators $\{\xi^\dag_{k\sigma}\}$ and $\{\eta^\dag_{k\sigma}\}$ are linear combinations of the original Jordan-Wigner fermion operators, we are allowed to use Wick's theorem to write the above expectation value as the Pfaffian of a $2(m+n)\times 2(m+n)$ antisymmetric matrix $A^{(m,n)}$~\cite{Fubini1952,Lieb1994}
\begin{eqnarray}\label{Ipf}
I_{m,n}&=&\frac{(-1)^m}{\prod^m_{l=1}v^{(\sigma)*}_{p_l}\prod^n_{j=1}v^{(\bar{\sigma})}_{k_j}} \mathrm{Pf}A^{(m,n)}.
\end{eqnarray}
For example, for $m=n=1$ we have (let $\langle\cdots\rangle_{\mathrm{v}}=\langle\mathrm{vac}|\cdots|\mathrm{vac}\rangle$)
\begin{eqnarray}\label{Ipf11}
&&\langle \xi_{p_1,\sigma}\eta_{p_1,\sigma} \xi^\dag_{k_1,\bar{\sigma}}\eta^\dag_{k_1,\bar{\sigma}} \rangle_{\mathrm{v}}\nonumber\\
&=&\langle \xi_{p_1,\sigma}\eta_{p_1,\sigma}\rangle_{\mathrm{v}}  \langle \xi^\dag_{k_1,\bar{\sigma}}\eta^\dag_{k_1,\bar{\sigma}} \rangle_{\mathrm{v}} -\langle  \xi_{p_1,\sigma}\xi^\dag_{k_1,\bar{\sigma}} \rangle_{\mathrm{v}} \langle \eta_{p_1,\sigma} \eta^\dag_{k_1,\bar{\sigma}} \rangle_{\mathrm{v}}\nonumber\\
&&+\langle \xi_{p_1,\sigma}\eta^\dag_{k_1,\bar{\sigma}} \rangle_{\mathrm{v}} \langle  \eta_{p_1,\sigma} \xi^\dag_{k_1,\bar{\sigma}} \rangle_{\mathrm{v}}\nonumber\\
&=&\mathrm{Pf}\left(
                \begin{array}{cccc}
                  0 & \langle \xi_{p_1,\sigma}\eta_{p_1,\sigma}\rangle_{\mathrm{v}} & \langle \xi_{p_1,\sigma}\xi^\dag_{k_1,\bar{\sigma}}\rangle_{\mathrm{v}} & \langle \xi_{p_1,\sigma}\eta^\dag_{k_1,\bar{\sigma}}\rangle_{\mathrm{v}} \\
                    & 0 & \langle \eta_{p_1,\sigma}\xi^\dag_{k_1,\bar{\sigma}} \rangle_{\mathrm{v}} & \langle  \eta_{p_1,\sigma}\eta^\dag_{k_1,\bar{\sigma}} \rangle_{\mathrm{v}} \\
                    & \ddots  & 0 &  \langle  \xi^\dag_{k_1,\bar{\sigma}}\eta^\dag_{k_1,\bar{\sigma}} \rangle_{\mathrm{v}} \\
                    &   &   & 0 \\
                \end{array}
              \right).\nonumber
\end{eqnarray}
To find out the nonvanishing entries of $A^{(m,n)}$, we note that there are six types of nonvanishing contractions
\begin{eqnarray}
\langle\mathrm{vac}|\xi_{p_l,\sigma}\eta_{p_l,\sigma}|\mathrm{vac}\rangle&=&-u^{(\sigma)*}_{p_l}v^{(\sigma)*}_{p_l},\nonumber\\
\langle\mathrm{vac}|\xi^\dag_{k_j,\bar{\sigma}}\eta^\dag_{k_j,\bar{\sigma}}|\mathrm{vac}\rangle&=& u^{(\bar{\sigma}) }_{k_j}v^{(\bar{\sigma}) }_{k_j},\nonumber\\
\langle\mathrm{vac}|\xi_{p_l,\sigma}\xi^\dag_{k_j,\bar{\sigma}}|\mathrm{vac}\rangle&=& v^{(\sigma)*}_{p_l}v^{(\bar{\sigma})}_{k_j} f_{-p_l,-k_j}\nonumber\\
\langle\mathrm{vac}|\xi_{p_l,\sigma}\eta^\dag_{k_j,\bar{\sigma}}|\mathrm{vac}\rangle&=&- v^{(\sigma)*}_{p_l}v^{(\bar{\sigma})}_{k_j} f_{-p_l,k_j}\nonumber\\
\langle\mathrm{vac}|\eta_{p_l,\sigma}\xi^\dag_{k_j,\bar{\sigma}}|\mathrm{vac}\rangle&=&- v^{(\sigma)*}_{p_l}v^{(\bar{\sigma})}_{k_j} f_{p_l,-k_j}\nonumber\\
\langle\mathrm{vac}|\eta_{p_l,\sigma}\eta^\dag_{k_j,\bar{\sigma}}|\mathrm{vac}\rangle&=& v^{(\sigma)*}_{p_l}v^{(\bar{\sigma})}_{k_j} f_{p_l, k_j},
\end{eqnarray}
where $f_{p,k}$ is given by Eq.~(\ref{fkkp}). Thus, the matrix $A^{(m,n)}$ has the following nonvanishing entries (for $l=1,2,\cdots,m$ and $j=1,2,\cdots,n$)
\begin{eqnarray}\label{Amnnonv}
A^{(m,n)}_{2l-1,2l}&=&-u^{(\sigma)*}_{p_l}v^{(\sigma)*}_{p_l},\nonumber\\
A^{(m,n)}_{2m+2j-1,2m+2j}&=&u^{(\bar{\sigma}) }_{k_j}v^{(\bar{\sigma}) }_{k_j},\nonumber\\
A^{(m,n)}_{2l-1,2m+2j-1}&=&v^{(\sigma)*}_{p_l}v^{(\bar{\sigma})}_{k_j} f_{-p_l,-k_j},\nonumber\\
A^{(m,n)}_{2l-1,2m+2j}&=&- v^{(\sigma)*}_{p_l}v^{(\bar{\sigma})}_{k_j} f_{-p_l,k_j},\nonumber\\
A^{(m,n)}_{2l,2m+2j-1}&=&- v^{(\sigma)*}_{p_l}v^{(\bar{\sigma})}_{k_j} f_{p_l,-k_j},\nonumber\\
A^{(m,n)}_{2l,2m+2j}&=&v^{(\sigma)*}_{p_l}v^{(\bar{\sigma})}_{k_j} f_{p_l, k_j}.
\end{eqnarray}
The matrix elements $A^{(m,n)}_{i,j}$ with $i>j$ can be obtained from the relation $A^{(m,n)}_{i,j}=-A^{(m,n)}_{j,i}$.
\par Fortunately, efficient numerical computation of the Pfaffian of $A^{(m,n)}$ can be achieved by using the software package presented in \cite{Wimmer}. Thus, $M_{x/y}(t)$ can be numerically calculated by combining Eqs.~(\ref{Mxy}), (\ref{lengthy}), (\ref{Ipf}), and (\ref{Amnnonv}). The dynamics of the transverse magnetization $M_z(t)=\langle  \psi^{\mathrm{(r)}}(t) |\sum^N_{j=1}\sigma^{z}_j|\psi^{\mathrm{(r)}}(t)\rangle$ can be calculated by noting that
\begin{eqnarray}\label{sumsigmaz}
&&\sum_i\sigma^z_i= (|-\pi\rangle \langle  -\pi|-|\mathrm{vac}\rangle_{-\pi }~_{-\pi }\langle \mathrm{vac}|)\nonumber\\
&&+(|0\rangle \langle 0|-|\mathrm{vac}\rangle_{0 }~_{0 }\langle \mathrm{vac}|)\nonumber\\
&&+ 2\sum_{k>0,k\in K_+}(|k,-k\rangle_+~_+\langle k,-k|-|\mathrm{vac}\rangle_{k,+}~_{k,+}\langle \mathrm{vac}|) \nonumber\\
          &&+2\sum_{k>0,k\in K_-}(|k,-k\rangle_-~_-\langle k,-k|-|\mathrm{vac}\rangle_{k,-}~_{k,-}\langle \mathrm{vac}|).\nonumber
\end{eqnarray}
Since $\sum_i\sigma^z_i$ preserves the fermion parity, $M_z(t)$ can be easily calculated as
\begin{eqnarray}
M_z(t)&=&\sum_{k>0,k\in K_+}\left[|v^{(+)}_k(t)|^2-|u^{(+)}_k(t)|^2\right]\nonumber\\
&&+\sum_{k>0,k\in K_-}\left[|v^{(-)}_k(t)|^2-|u^{(-)}_k(t)|^2\right].
\end{eqnarray}
\subsection{Numerical examples}
\par In this subsection, we apply the above formalism for calculating the time-evolution of the longitudinal magnetizations to two physical scenarios, namely sudden quenches in the magnetic field and periodic delta kicks. In both cases the driving Hamiltonians are translationally invariant so that the Pfaffian method developed in the last subsection can be utilized.
%\par In the next subsection we will discuss several examples for which the procedure developed in this subsection can be applicable.
\subsubsection{Sudden quench}
\par Our first example is the time evolution governed by the Hamiltonian of the quantum Ising ring with finite magnetic field $g$. This dynamical process can be viewed as a sudden quench of $g$ from the initial value $g_i=0$ to the final value $g_f>0$. In this case the driving Hamiltonian is consistent with Eq.~(\ref{mHe}), i.e., $\mathcal{H}^{(\mathrm{driv})}_{k,\sigma}=\mathcal{H}^{(\mathrm{e})}_{k,\sigma}$ and $\gamma(t)=\frac{1}{2}\int^t_0ds[-2(1-g_f)-2(1+g_f)]=-2t$.
\begin{figure}
\includegraphics[width=.48\textwidth]{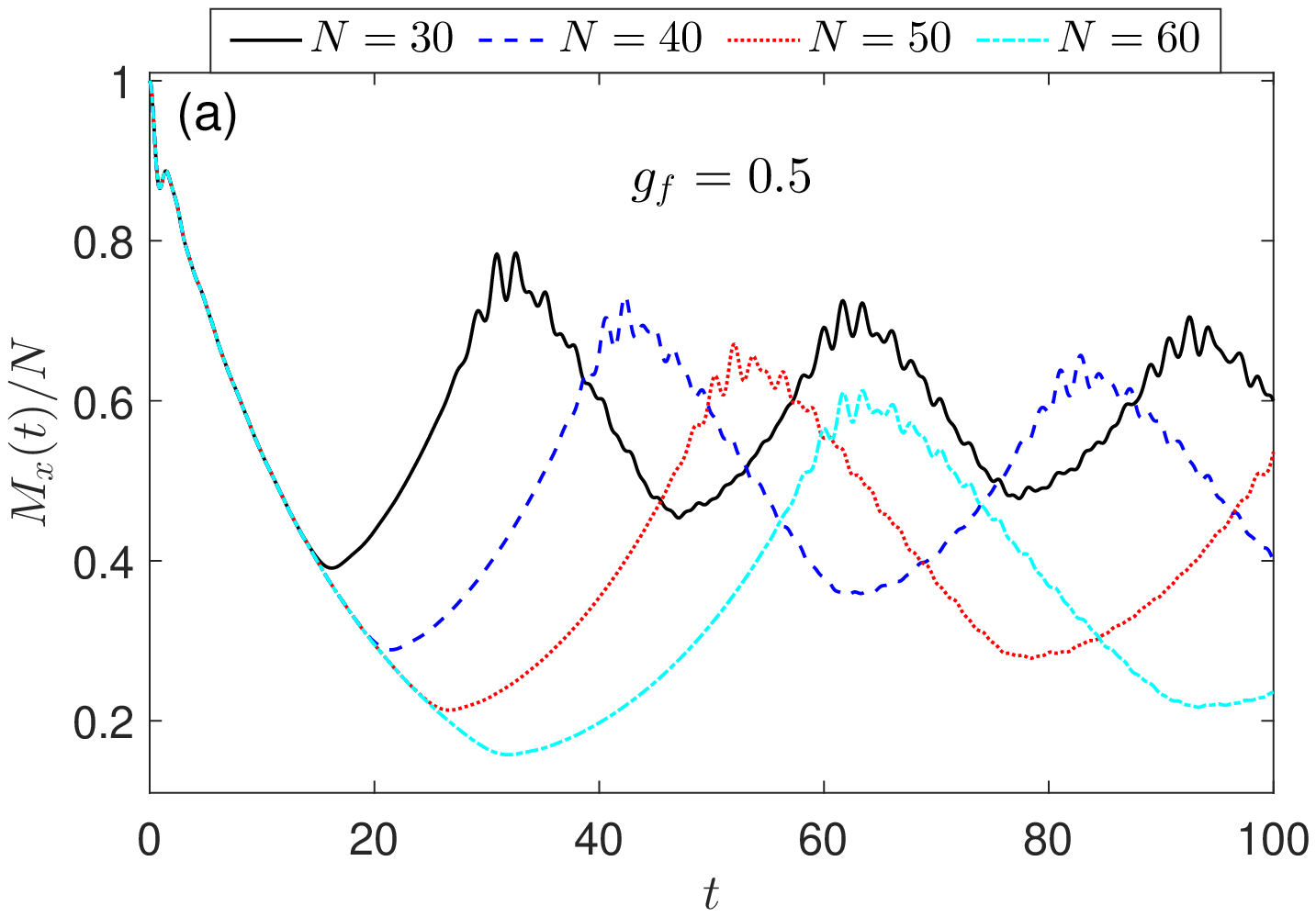}
\includegraphics[width=.48\textwidth]{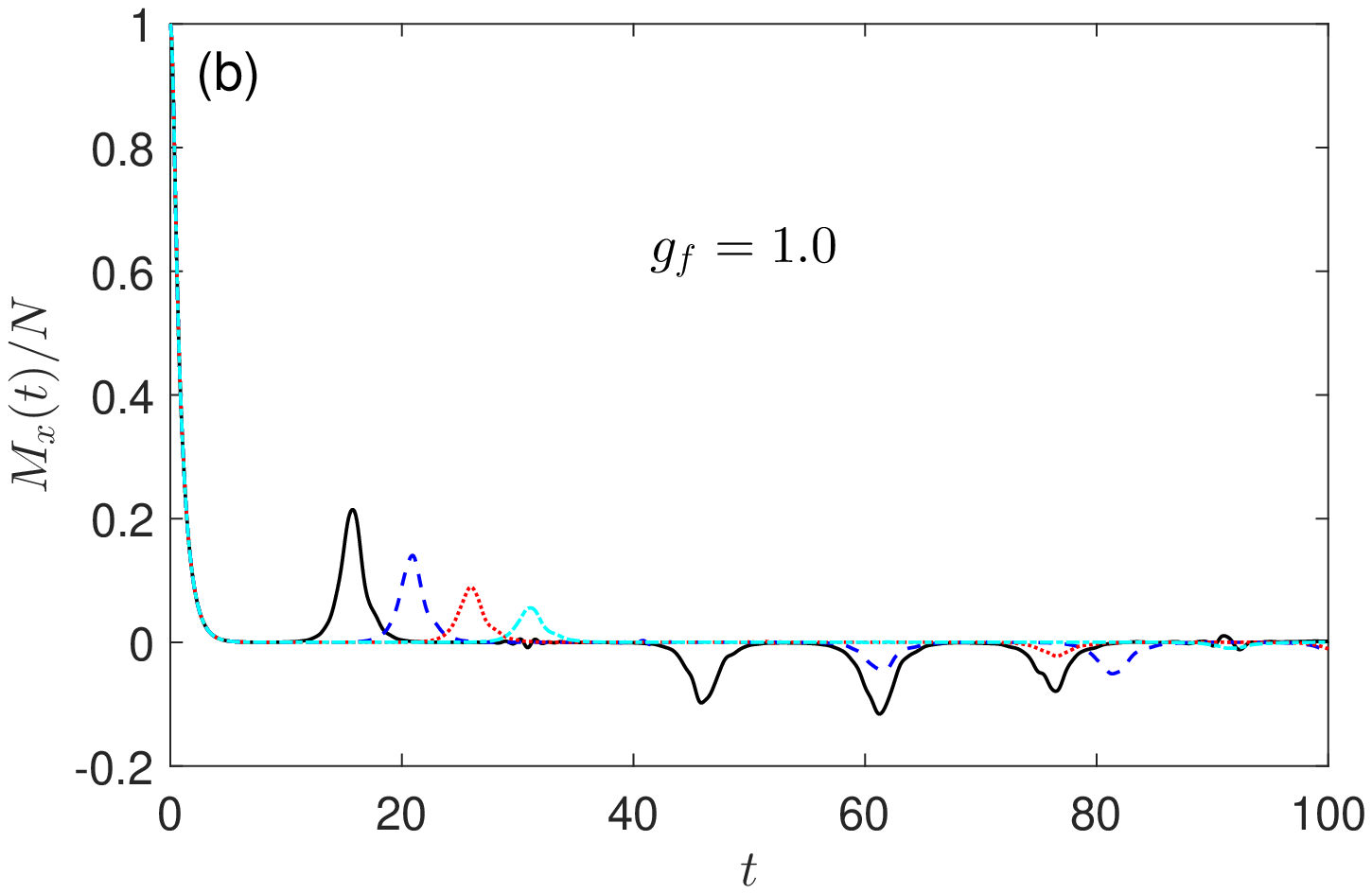}
\includegraphics[width=.48\textwidth]{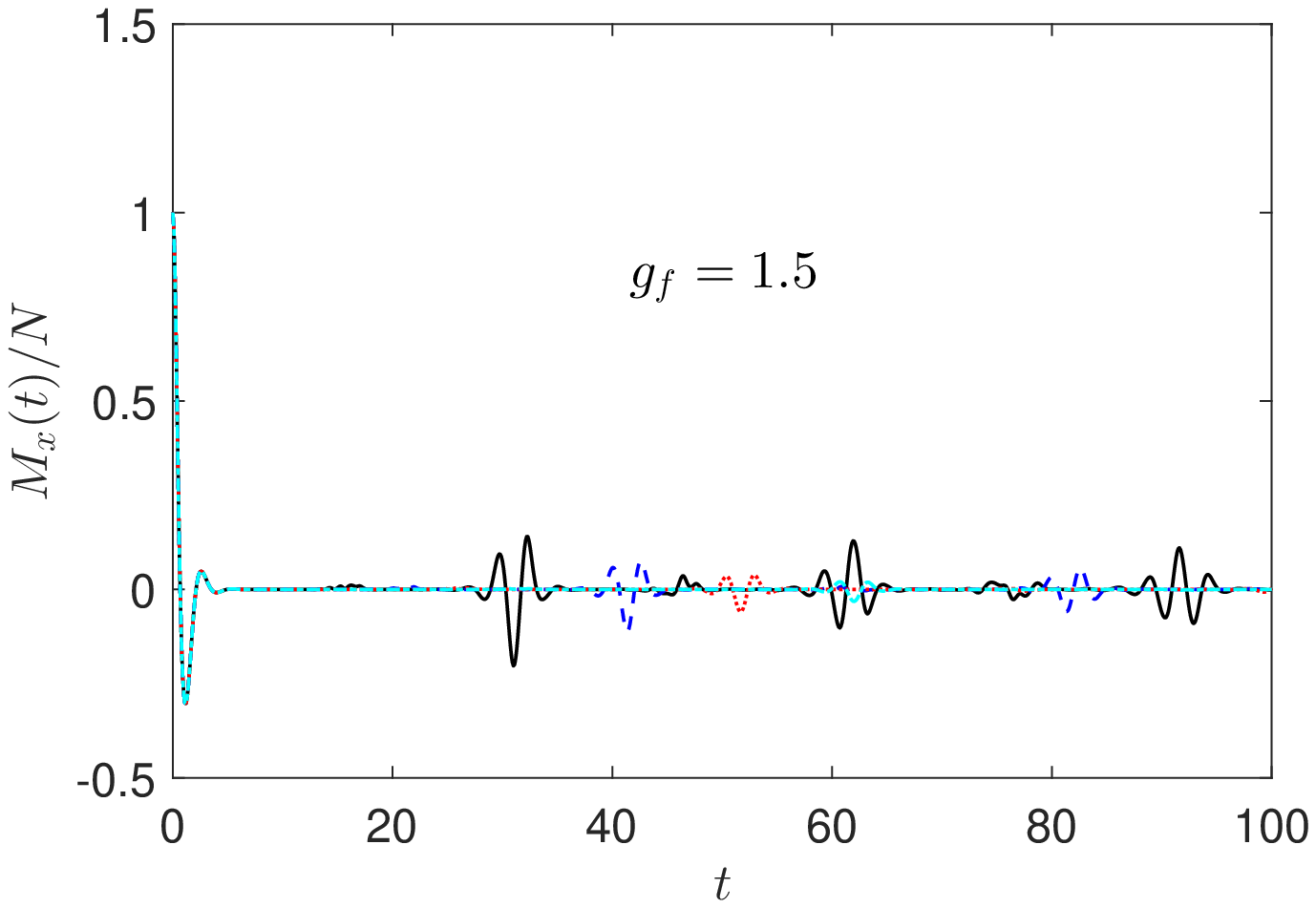}
\caption{Dynamics of the longitudinal magnetization $M_x(t)/N$ after a sudden quench of the magnetic field from $g_i=0$ to $g_f$. (a) $g_f=0.5$, (b) $g_f=1.0$, (c) $g_f=1.5$. The initial state is prepared as the ferromagnetic state $|\psi^{\mathrm{(r)}}_{g=0}\rangle=|\rightarrow\rangle_1\cdots|\rightarrow\rangle_N$.}
\label{Fig3}
\end{figure}
\par Figure~\ref{Fig3}(a) shows the evolution of $M_x(t)/N$ for a sudden quench to $g_f=0.5$, which occurs within the ordered phase. Results for different system sizes $N=30$, $40$, $50$, and $60$ are presented for comparison. It can be seen that $M_x(t)/N$ experiences a universal decay that is almost independent of the system size $N$ at short-times scales. At a time around $t_{\min}\approx N/2$, $M_x(t)/N$ reaches its first minimum, $M_x(t_{\min})/N$, which decreases with increasing $N$. Numerical fittings approximately give $t_{\min}=0.525 N+0.4$ and $M^{(\mathrm{\min})}_x(t)/N=0.9744 e^{-0.0304 N}$, suggesting that $M_x(t_{\min})$ decreases exponentially with increasing $N$. After that, $M_x(t)/N$ exhibits a quasi-periodic oscillatory behavior with period $T\approx N$. It is expected that $M_x(t)/N$ decays exponentially in the thermodynamics limit $N\to\infty$ and approaches zero in the long-time limit. These observations are qualitatively consistent with previous works which deal with quenches within the ordered phase in either infinite systems~\cite{calab2001} or for open boundary conditions~\cite{Igloi2011PRL,Igloi2011PRB}.
\par Figure~\ref{Fig3}(b) shows the dynamics of $M_x(t)/N$ for a sudden quench to the critical point $g_f=1.0$. A universal but abrupt short-time decay to nearly zero is observed. In contrast to the quench within the critical phase, $M_x(t)/N$ reaches its first maximum around $t_{\max}\approx N/2$. These maxima seem decrease exponentially with increasing $N$. Numerical fittings approximately give $t_{\max}=0.515 N+0.24$ and $M_x(t_{\max})/N=0.8316 e^{-0.04478 N}$. For long enough chains, the magnitude of $M_x(t)/N$ tends toward zero after the initial decay process.
\par Figure~\ref{Fig3}(c) shows the dynamics of $M_x(t)/N$ for a sudden quench to the disordered phase with $g_f=1.5$. We observe more abrupt short-time decay of $M_x(t)/N$ to negative values, but followed by a sharp increase to nearly zero. After that the magnitude of $M_x(t)/N$ is nearly vanishing but exhibits minor revivals at around $t\approx mN,~m\in Z$.
\subsubsection{delta kick}
\par The second example we will discuss is the periodic delta kick studied in~\cite{Yu2019}. Following Ref.~\cite{Yu2019}, we consider the following protocol. Suppose we start with $|\psi(0)\rangle=|\psi^{\mathrm{(r)}}_{g=0}\rangle$ and evolve the system with the quantum Ising Hamiltonian $H_{\mathrm{QIM}}(g)$ with fixed magnetic field $g$ for a time $\tau$. We then apply a delta kick
\begin{eqnarray}
K_\phi=e^{-i\frac{\phi}{2}\sum_i\sigma^z_i},
\end{eqnarray}
which rotates the spins about the $z$ axis by an angle
\begin{eqnarray}
\phi=\pi\left(1- \epsilon\right),
\end{eqnarray}
where $\epsilon$ is a perturbation. The time-evolution operator over one period is
\begin{eqnarray}
U(\tau)=K_{\pi\left(1- \epsilon\right)}e^{-iH_{\mathrm{QIM}}(g)\tau}.
\end{eqnarray}
The state of the system just after the $n$th kick is thus
\begin{eqnarray}
|\psi(n)\rangle&=&U^n|\psi(0)\rangle.
\end{eqnarray}
For $\epsilon=0$ and $g=0$, the initial state is an eigenstate of $H_{\mathrm{QIM}}(g=0)$, so that the spins flip at $n\tau$ and the system returns to the initial state at every $2\tau$.
\par Note that both $H_{\mathrm{QIM}}$ and $K_\phi$ conserve the parity of the fermions and are translationally invariant, so we can write
\begin{eqnarray}
&&e^{-iH_{\mathrm{QIM}}(g)\tau}\nonumber\\
&=&P_-e^{-\frac{1}{2}i\mathcal{H}_{-\pi}\tau} e^{-\frac{1}{2}i\mathcal{H}_{0}\tau}\prod_{k>0,k\in K_-}e^{-i\mathcal{H}^{\mathrm{(e)}}_{k,-}\tau}P_-\nonumber\\
&&+P_+\prod_{k>0,k\in K_+}e^{-i\mathcal{H}^{\mathrm{(e)}}_{k,+}\tau}P_+,
\end{eqnarray}
and
\begin{eqnarray}\label{KphiF}
K_\phi&=& P_-e^{-i\frac{\phi}{2}\mathcal{F}_{-\pi}}e^{-i\frac{\phi}{2}\mathcal{F}_{0}}\prod_{k>0,k\in K_-}e^{-i\frac{\phi}{2}\mathcal{F}_{k,-}} P_-\nonumber\\
&&+P_+\prod_{k>0,k\in K_+}e^{-i\frac{\phi}{2}\mathcal{F}_{k,+}}  P_+,
\end{eqnarray}
\begin{figure}
\includegraphics[width=.50\textwidth]{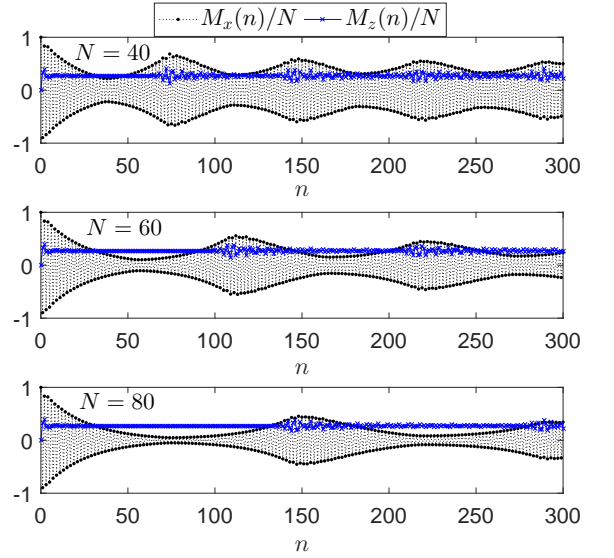}
\caption{Stroboscopic longitudinal magnetization $M_x(n)/N$ and transverse magnetization $M_z(n)/N$ as functions of number of kicks $n$ for various system sizes $N$. The initial state is prepared as the ferromagnetic state $|\psi^{\mathrm{(r)}}_{g=0}\rangle=|\rightarrow\rangle_1\cdots|\rightarrow\rangle_N$. Parameters: $g=0.5,~\tau=0.5$, and $\epsilon=0.02$.}
\label{Fig4}
\end{figure}
where
\begin{eqnarray}
\mathcal{F}_{k,\sigma}&=&2\left(
                          \begin{array}{cc}
                            -1 & 0 \\
                            0 & 1 \\
                          \end{array}
                        \right),\nonumber\\
                      ~\mathcal{F}_{-\pi}&=&\left(
                          \begin{array}{cc}
                            -1 & 0 \\
                            0 & 1 \\
                          \end{array}
                        \right),~\mathcal{F}_{0}=\left(
                          \begin{array}{cc}
                            -1 & 0 \\
                            0 & 1 \\
                          \end{array}
                        \right).
\end{eqnarray}
We have used Eq.~(\ref{sumsigmaz}) in deriving Eq.~(\ref{KphiF}).
Hence,
\begin{eqnarray}
&&U^n=P_+\prod_{k>0,k\in K_+}\left(e^{-i\frac{\phi}{2}\mathcal{F}_{k,+}}e^{-i\mathcal{H}^{\mathrm{(e)}}_{k,+}\tau}\right)^n  P_+\nonumber\\
&&+P_-\left(e^{-i\frac{\phi}{2}\mathcal{F}_{-\pi}}e^{-\frac{1}{2}i\mathcal{H}_{-\pi}\tau}\right)^n\left(e^{-i\frac{\phi}{2}\mathcal{F}_{0}}e^{-\frac{1}{2}i\mathcal{H}_{0}\tau}\right)^n\nonumber\\
&&\prod_{k>0,k\in K_-}\left(e^{-i\frac{\phi}{2}\mathcal{F}_{k,-}}e^{-i\mathcal{H}^{\mathrm{(e)}}_{k,-}\tau}\right)^n P_-.
\end{eqnarray}
When acting on the initial state, we finally get
\begin{eqnarray}
&&|\psi(n)\rangle=\frac{1}{\sqrt{2}} \prod_{k>0,k\in K_+}\left(e^{-i\frac{\phi}{2}\mathcal{F}_{k,+}}e^{-i\mathcal{H}^{\mathrm{(e)}}_{k,+}\tau}\right)^n  \chi_{k,+}\nonumber\\
&&+\frac{e^{-i\frac{\pi}{4}}e^{i2n\tau}}{\sqrt{2}}\chi_{-\pi}\chi_{0} \nonumber\\
&&\prod_{k>0,k\in K_-}\left(e^{-i\frac{\phi}{2}\mathcal{F}_{k,-}}e^{-i\mathcal{H}^{\mathrm{(e)}}_{k,-}\tau}\right)^n \chi_{k,-},
\end{eqnarray}
where
\begin{eqnarray}
\chi_{k,\sigma}=\left(
             \begin{array}{c}
               \sin\frac{k}{2} \\
               \cos\frac{k}{2} \\
             \end{array}
           \right),~\chi_{-\pi}=\left(
             \begin{array}{c}
              1 \\
               0 \\
             \end{array}
           \right),~\chi_{0}=\left(
             \begin{array}{c}
             0 \\
              1 \\
             \end{array}
           \right).\nonumber\\
\end{eqnarray}
The expectation value of the longitudinal magnetization after the $n$th kick, $M_x(n)=N\langle\psi(n)|\sum_i\sigma^x_i|\psi(n)\rangle$, can now be calculated by using the Pfaffian method presented in Sec.~\ref{Generalform}.
\par In Fig.~\ref{Fig4} we show both the stroboscopic longitudinal magnetization $M_x(n)/N$ and transverse magnetization $M_z(n)/N$ as functions of the number of kicks $n$ for different system sizes. It can be seen that $M_x(n)/N$ breaks the time-translational symmetry but shows a persistent oscillatory profile with a fixed period approximately proportional to $N$. While the transverse magnetization $M_z(n)/N$ only experiences small oscillators when $M_x(n)/N$ showing peaks. These results are consistent with those obtained in~\cite{Yu2019}. However, the numerical simulations performed here are far beyond the reach of exact diagonalization used in Ref.~\cite{Yu2019}. Our formalism thus provides a numerically efficient way to study the emergence of discrete time crystals in the quantum Ising setup with finite but large number of spins.
\section{Conclusions}\label{V}
\par In this work, we revisit several aspects of the finite-size quantum Ising chain with periodic boundary conditions. In special, we show that the momentum-space BCS-type ground states of the classical Ising ring are proportional to the equally weighted linear superpositions of the two fully aligned ferromagnetic states. This relationship is a special case of the correspondence between the spatially factorized ground states of the XYZ spin chain under frustration-free conditions and the two fermionic states with distinct fermion parities in the Jordan-Wigner fermion representation.
\par Based on the above relationship between the real-space and momentum-space representations of the same state, we study the real-time dynamics of the longitudinal magnetization $\sum_i\sigma^x_i$ under translationally invariant driving Hamiltonians, with the system prepared in one of the ferromagnetic states. Since the ferromagnetic state is a linear superposition of states with distinct fermion parities, the calculation of the parity-breaking longitudinal magnetization dynamics is less straightforward. Fortunately, by writing the BCS mode state as the occupation state of two time-dependent effective Bogoliubov fermions, we are able to drive a Pfaffian formula for the calculation of the longitudinal magnetization dynamics. The obtained formalism is then applied to two dynamical scenarios, namely the sudden quench and delta kicks. With the help of software package that can realize efficient numerical computation of Pfaffians, we perform numerical simulations for both scenarios in relatively large systems.

\par \emph{Note added}. While the present work was nearly finished, we became aware of a related work~\cite{Damskiarxiv} in which the authors also studied the dynamics of the longitudinal magnetization starting with a ferromagnetic state. Though there is certain amount of overlap between the two works, we note that in Ref.~\cite{Damskiarxiv} the dynamics of the longitudinal magnetization in a periodic chain, which is the main focus of our work, is calculated for at most $N=12$ spins using exact diagonalization.

\noindent{\bf Acknowledgements:}
We thank Hosho Katsura for useful discussions on the Perron-Frobenius theorem. We also thank Wen-Long You for bringing Ref.~\cite{Yu2019} to our attention. This work was supported by the Natural Science Foundation of China (NSFC) under Grant No. 11705007, and partially by the Beijing Institute of Technology Research Fund Program for Young Scholars.

\appendix
\section{Determination of the ground-state fermion parity using the Perron-Frobenius theorem}\label{app1}
\par We closely follow Ref.~\cite{Katsura2012} to show that the ground state $|\psi_{\mathrm{G}}\rangle$ of $H_{\mathrm{QIM}}(g)$ with even (odd) number of sites has an even (odd) fermion parity for $g>0$. Using the basis in which $\sigma^x_j$ is diagonalized, i.e., $|\rightarrow\rangle_j=(|\uparrow\rangle_j+|\downarrow\rangle_j)/\sqrt{2}$, $|\leftarrow\rangle_j=(|\uparrow\rangle_j-|\downarrow\rangle_j)/\sqrt{2}$, all of the off-diagonal elements of $H_{\mathrm{QIM}}$ are nonpositive and satisfy the connectivity condition (note that $\sigma^z_j|\rightarrow/\leftarrow\rangle_j=|\leftarrow/\rightarrow\rangle_j$). According to the Perron-Frobenius theorem, the ground state of $H_{\mathrm{QIM}}$ is nondegenerate and has the form
\begin{eqnarray}\label{psiG}
|\psi_{\mathrm{G}}\rangle=\sum_{\alpha_1,\cdots,\alpha_N=\pm1}C_{\alpha_1,\cdots,\alpha_N}\prod^N_{j=1}(|\uparrow\rangle_j+\alpha_j|\downarrow\rangle_j),
\end{eqnarray}
where the coefficients $C_{\alpha_1,\cdots,\alpha_N}$ are strictly positive for any $(\alpha_1,\cdots,\alpha_N)$. Due to the $Z_2$ symmetry of $H_{\mathrm{QIM}}$ under $\sigma^x_j\to-\sigma^x_j$, the coefficients $C_{\alpha_1,\cdots,\alpha_N}$ also satisfy
\begin{eqnarray}\label{CC}
C_{\alpha_1,\cdots,\alpha_N}=C_{-\alpha_1,\cdots,-\alpha_N}.
\end{eqnarray}
Hence, $\prod^N_{j=1}(|\uparrow\rangle_j+\alpha_j|\downarrow\rangle_j)$ and $\prod^N_{j=1}(|\uparrow\rangle_j-\alpha_j|\downarrow\rangle_j)$ always appear in pairs with equal coefficients. By noting that $|\uparrow\rangle_j$ ($|\downarrow\rangle_j$) has an odd (even) fermion parity in the fermion representation, we immediately see that $|\psi_{\mathrm{G}}\rangle$ is an even (odd) state for even (odd) $N$.
\section{Proof of inequality (\ref{DeltalP}) for $OP\geq1$}\label{app2}
\par Note that $f_{N}(x)=x+1$, so $\Delta l(P)$ given by Eq.~(\ref{PAPB1}) can be rewritten as
\begin{eqnarray}
\Delta l(P)&=&\sum^{N/2}_{j=1}[f_{2j-1}(x) - f_{2j}(x)] +x+1\nonumber\\
&=&2x\sum^{N/2}_{j=1}\frac{\cos2j\alpha-\cos(2j-1)\alpha}{f_{2j-1}(x) + f_{2j}(x)} +x+1.\nonumber\\
\end{eqnarray}
We now observe that $\{f_j(x)\}$ are all monotonically increasing functions of $x$ on $x\in[\cos\alpha,+\infty)$ and $\cos2j\alpha<\cos(2j-1)\alpha$ for all $j$, which implies that $\Delta l(P)$ is also monotonically increasing on $x\in[\cos\alpha,+\infty)$. Moreover, for $P=Q=(0,1)$ we have $\Delta l(Q)=1+\tan\frac{\pi}{4N}>1$. We thus get
\begin{eqnarray}
\Delta l(P)>\Delta l(Q)>1,~(OP\geq1),
\end{eqnarray}
which proves the inequality (\ref{DeltalP}) for $x\geq1$.
\section{Factorized ground states of the XYZ spin chain under the frustration-free condition}\label{app3}
\par We review the frustration-free condition for the XYZ spin chain under which factorized ground states are admitted. Consider the XYZ spin chain with open boundary conditions
\begin{eqnarray}
H^{(\mathrm{OBC})}_{\mathrm{XYZ}}&=&\frac{1}{4}\sum^{N-1}_{j=1}(J_x\sigma^x_j\sigma^x_{j+1}+J_y\sigma^y_j\sigma^y_{j+1}+J_z\sigma^z_j\sigma^z_{j+1})\nonumber\\
&&-\frac{1}{2}\sum^{N}_{j=1}h_j\sigma^z_j,
\end{eqnarray}
where we assume $J_x<J_y\leq0$ and $J_z,h_j\geq 0$ in order to incorporate the Hamiltonian of the quantum Ising model, and the inhomogeneous magnetic fields are chosen as $h_1=h_N=h/2$ and $h_2=h_3=\cdots=h_{N-1}=h$. It is known that~\cite{Krumann,Muller1985,Cerezo2015,Katsura2015} $H^{(\mathrm{OBC})}_{\mathrm{XYZ}}$ possesses two degenerate spatially separable ground states on the frustration-free hypersurface defined by
\begin{eqnarray}\label{hstar}
h&=&h^*,
\end{eqnarray}
where
\begin{eqnarray}
h^*&\equiv&\sqrt{(J_z-J_x)(J_z-J_y)}.
\end{eqnarray}
The two ground states are of the form
\begin{eqnarray}
&&|\Psi^{(\pm)}\rangle=\frac{1}{(1+\beta^*)^{N/2} }\nonumber\\
&&(|\downarrow\rangle_1\pm\sqrt{\beta^*}|\uparrow\rangle_1)\cdots(|\downarrow\rangle_N\pm\sqrt{\beta^*}|\uparrow\rangle_N),\nonumber\\
\end{eqnarray}
where
\begin{eqnarray}\label{cotthetastar}
\beta^*= \frac{-(J_x-J_y)}{\sqrt{(J_x-J_y)^2+4h^{*2}}-2h^*}>0.
\end{eqnarray}
It is important to note that
\begin{eqnarray}
\langle \Psi^{(+)} |\Psi^{(-)}  \rangle&=&\langle \Psi^{(-)} |\Psi^{(+)}  \rangle=\left(\frac{1-\beta^*}{1+\beta^*}\right)^N,
\end{eqnarray}
which means $|\Psi^{(+)}  \rangle$ and $|\Psi^{(-)}  \rangle$ are nonorthogonal unless $\beta^*=1$, or $h^*(J_x-J_y)=0$. This gives the condition
\begin{eqnarray}
h^*=0
\end{eqnarray}
provided $J_x\neq J_y$. Since we already assumed $J_x<J_y\leq 0$ and $J_z\geq 0$, it is easy to see from Eq.~(\ref{hstar}) that the only possibility is $J_z=J_y=0$, for which $H^{(\mathrm{OBC})}_{\mathrm{XYZ}}$ reduces to the classical ferromagnetic Ising chain with free ends.
\par The equally weighted superpositions of $|\Psi^{(+)}\rangle$ and $|\Psi^{(-)}\rangle$ form two orthogonal ground states
\begin{eqnarray}
|\tilde{\Psi}^{(\mathrm{e})}\rangle&=&\frac{1}{\sqrt{2}}(|\Psi^{(+)}\rangle+|\Psi^{(-)}\rangle),\nonumber\\
|\tilde{\Psi}^{(\mathrm{o})}\rangle&=&\frac{1}{\sqrt{2}}(|\Psi^{(+)}\rangle-|\Psi^{(-)}\rangle),
\end{eqnarray}
which have distinct fermion parities when expressed in terms of the Jordan-Wigener fermions. Actually, it was shown in Ref.~\cite{Katsura2017} that $|\tilde{\Psi}^{(\mathrm{e})}\rangle$ ($|\tilde{\Psi}^{(\mathrm{o})}\rangle$) is exactly the ground state of the \emph{fermion version} of the homogeneous \emph{periodic} XYZ chain
\begin{eqnarray}
H^{(\mathrm{PBC})}_{\mathrm{XYZ}}&=&\frac{1}{4}\sum^{N}_{j=1}(J_x\sigma^x_j\sigma^x_{j+1}+J_y\sigma^y_j\sigma^y_{j+1}+J_z\sigma^z_j\sigma^z_{j+1})\nonumber\\
&&-\frac{h}{2}\sum^{N}_{j=1}\sigma^z_j,
\end{eqnarray}
in the even (odd) parity sector after the Jordan-Wigner transformation.
\par In the case of $J_x=-4,~J_y=J_z=0$ and $h=2g$, the Hamiltonian $H^{(\mathrm{PBC})}_{\mathrm{XYZ}}$ reduces to the Hamiltonian of the quantum Ising ring given by Eq.~(\ref{HQIM}). From Eqs.~(\ref{hstar}) and (\ref{cotthetastar}), we have $h^*=0$ and $\beta^*=1$, so that the two states $|\Psi^{(\pm)}\rangle$ are orthogonal and reduce to $|\psi^{\mathrm{(r/l)}}_{g=0}\rangle$ given by Eqs.~(\ref{FMright}) and (\ref{FMleft}).

\end{document}